# Statistical analysis of the material, geometrical and imperfection characteristics of structural stainless steels and members

Itsaso Arrayago[a,*], Kim J.R. Rasmussen[b], Esther Real[a]

[a] *Universitat Politècnica de Catalunya, Dept. of Civil and Environmental Engineering, Spain*

[b] *The University of Sydney, School of Civil Engineering, Australia*

* Corresp.author: Itsaso Arrayago, Jordi Girona 1-3, Barcelona, Spain, itsaso.arrayago@upc.edu

**ABSTRACT**

Traditional member-based *two-step* design approaches included in current structural codes for steel structures, as well as more recent system-based *direct-design* alternatives, require building rigorous structural reliability frameworks for the calibration of partial coefficients (resistance factors) to achieve specified target reliability requirements. Key design parameters affecting the strength of structures and their random variations are generally modelled by nominal or characteristic values in design standards, which are combined with partial coefficients that need to be calibrated from measurements on real samples. While the statistical characterization of material and geometric properties of structural steels has been consolidated over the last decades, information about the characterization of structural stainless steels is virtually non-existent due to the limited pool of available data. Thus, this paper presents the basic ingredient for developing reliability frameworks for stainless steel structures and components by statistically characterizing the main random parameters affecting their strength through a comprehensive database collected from the literature. Based on the collected data, appropriate probabilistic models are proposed for geometric properties, material properties, imperfections and residual stresses of different stainless steel alloys and cross-section or product types. The data is equally applicable to member-based reliability analyses as described in existing codes and system-based analyses targeted at the *direct-design* of stainless steel structures by advanced analysis.

**HIGHLIGHTS**

- Data on most relevant random variables is collected through a comprehensive literature review.

- Data for different stainless steel alloys and cross-section types are analysed.

- The variability of parameters affecting the strength of stainless steel structures is statistically characterized.

- Appropriate statistical functions and probabilistic models are provided for geometry, material, imperfections and residual stresses.

- Foundational data is provided for complete member- and system-based reliability frameworks, from which safety factors and resistance factors can be derived.

## 1. INTRODUCTION

Design of structures is based on traditional limit state criteria following a *two-step* approach where internal actions are first determined from a structural analysis and limit states are subsequently checked for members and connections. Provisions for these checks are given by structural codes for the particular materials, e.g. steel [1-4], in which design parameters and their random variations are modelled by nominal or characteristic values affected by partial coefficients (also referred to as resistance factors), which have been calibrated to achieve a specified target reliability requirement. For the calibration of these partial coefficients it is fundamental to have objective information on the variability of the different random parameters affecting the strength of real structures, which can be obtained and verified from measurements and tests on real samples.

Over the last decades extensive research works have been devoted to the statistical characterization of structural steels, mainly focused on the material (especially the yield stress and the ultimate strength) and geometric (flange/web thicknesses, web depths, flange width) properties of hot-rolled steels [5-11], although studies on variables such as the Young's modulus, ultimate strain and initial imperfections are considerably more limited. This information has been used for the calibration of resistance factors and as input data for a number of reliability studies using the Monte Carlo method [7,9] or for the development of direct system-based design-by-analysis approaches such as the Direct Design Method (DDM) [12-16]. With a considerably more limited pool of available data, equivalent studies providing a statistical characterization of structural stainless steels are scarce and limited to geometric variables and basic material properties (yield stress and ultimate strength) [17-19], and inexistent for other variables including key material properties such as the Young's modulus, ultimate strain and strain hardening exponents as well as initial geometric imperfections and residual stresses.

With the aim of providing the required input data for the calibration of resistance factors and other reliability studies, this paper presents the statistical characterisation of the variability and distribution functions of the main random parameters affecting the strength of stainless steel structures. The analysis is based on a comprehensive database collected from the literature for the most relevant random variables, including geometric properties, material properties, imperfections and residual stresses for the most common stainless steel cross-section types and alloys. For each random variable, appropriate statistical parameters as well as probabilistic models derived from the data are proposed, and correlations between the different variables are reported. The characterisation of the material, geometric and imperfection characteristics of stainless steel structures will serve as mandatory input for future member- and system-based reliability calibrations.

## 2. METHODOLOGY

This section presents a detailed description of the methodology employed in this paper for the assembly and statistical characterisation of the random variables defining geometric and material properties, imperfections and residual stresses of stainless steel structures, including information on the data collection, identification and fitting of the most adequate probabilistic distributions and the evaluation of the goodness-of-fit of the proposed distributions.

*Assembly of databases and descriptive statistics*

The databases utilized for the statistical characterisation of random variables have been primarily assembled from measurements reported in the literature by international research groups in the framework of different experimental programmes. The assembled databases include geometric properties (section height, width, thickness and corner radius), material parameters (Young's modulus, yield stress, ultimate tensile strength, ultimate strain and strain hardening exponents), imperfections (at local cross-section, member and frame levels) and residual stresses. The study assumes that the measured dimensions reported in the literature are representative of specimens delivered in practical applications. This assumption is premised on the fact that the experimental specimens in most of the reported cases were supplied by the main international stainless steel producers, which are considerably more limited in numbers than structural steel producers. In general, measured values of the random variables have been compared with the corresponding nominal values when analysing the data and proposing statistical distributions. Actual measurements of the different random variables investigated were directly extracted from the literature, while the corresponding nominal values were obtained from the papers or the relevant standards. For a limited number of cases, and especially for material parameters and residual stress magnitudes, raw data was directly obtained from the corresponding authors and processed in this study, as indicated in each of the relevant sections.

Independent datasets from different sources were assembled in a general database for each random variable, but sub-groups corresponding to each dataset were kept identifiable throughout the analysis. Once the complete datasets were collated, basic descriptive statistics were calculated for a preliminary analysis of each of the random variables, including the mean, standard deviation, coefficient of variation (COV), skewness and kurtosis. All parameters were calculated by means of the predefined functions included in the numerical computing software MATLAB R2020 [20] and the statistics package Minitab 18 [21]. These parameters, and in particular the skewness and kurtosis, contributed to the identification of the most suitable distribution type for each variable. The weighted mean and pooled standard deviations reported in this paper were calculated from Eq. 1 and Eq. 2 in order to minimize the effect of the different measurement and data management techniques used by independent authors. In these equations $n_i$ is the number of measurements in data set *i*, while $\mu_{X,i}$ and $\sigma_{X,i}$ are the mean value and standard deviation of the random variable X for data set *i*, and *k* is the total number of individual datasets considered. The random variable X is the ratio between a measured dimension and its nominal value. Note that the pooled standard deviation is an estimate of the population standard deviation calculated using the observations in more than one sample or dataset, and thus presumes that the standard deviations of the observations are the same for all the samples or individual datasets. Hence, it is assumed that the measurements reported by independent researchers stem from populations with common statistical distribution. Since for small sample sizes the standard error of the sample variance can be significantly large, it is preferable to pool data when possible to obtain a more precise estimate of the standard deviation [22].

$$\mu_{X,\text{weighted}} = \frac{\sum_{i=1}^{k} n_i \mu_{X,i}}{\sum_{i=1}^{k} n_i} \qquad \text{Eq. 1}$$

$$\sigma_{X,\text{pooled}} = \sqrt{\frac{\sum_{i=1}^{k}(\sigma_{X,i}^2 (n_i - 1))}{\sum_{i=1}^{k} n_i - k}} \qquad \text{Eq. 2}$$

*Selection and fitting of underlying distributions*

The most common probability distributions to model random variables related to geometric parameters and material properties found in the literature are the normal or log-normal distributions [5,6,23], while Extreme Type distributions are more commonly adopted for modelling load effects [24,25]. The choice of the underlying distributions for the random variables investigated in this paper (mainly geometric, material and imperfection parameters) was carried out from the histograms of the assembled data samples for the candidate distributions (i.e. normal or log-normal distributions) and assisted by previous experience for similar variables reported in the literature. Since the skewness is an indirect measure of the asymmetry of the data around the sample mean (with positive skewness values indicating that the data spreads out more to the right) and the kurtosis indicates the peakiness or how outlier-prone a distribution is (with high positive values indicating a sharp peak), these two parameters can be used as preliminary normality indicators (with the normal distribution showing skewness and kurtosis parameters close to zero and three, respectively) [22]. Although the final fit of the proposed distribution was based on the maximum likelihood method using MATLAB [20], graphical methods were employed as secondary aids.

*Evaluation of goodness-of-fit*

Before the most suitable probabilistic distribution was proposed, the sample data was compared with the resulting distributions through the use of goodness-of-fit tests, the most common of which are the Kolmogorov-Smirnov and the Anderson-Darling (AD) tests. Although both tests are based on empirical distribution functions and are suitable for identifying a distribution as not normal when it is skewed, the Anderson-Darling test tends to be more effective in detecting departures in the tails of the distribution, as it places greater weight on the observations in those regions, and thus it has been employed in this paper [22]. Anderson-Darling tests at the 5% significance level were carried out by means of the function included in the Statistics and Machine Learning Toolbox in MATLAB [20]. Finally, the goodness of the distributions was also assessed by visual diagnosis of the probability plots and the comparison of the empirical cumulative distribution functions (CDF) against the hypothesized cumulative distribution functions.

## 3. GEOMETRIC PROPERTIES: CROSS-SECTION DIMENSIONS

Cross-section dimensions are fundamental variables affecting the strength and stability of stainless steel structures. Current design provisions include design expressions that explicitly depend on magnitudes directly associated with nominal cross-section dimensions, such as cross-section area, second moment of area and section modulus, and are usually provided by producers. However, these nominal dimensions are slightly different from the measured actual dimensions, presenting random variability due to the manufacturing processes, and thus it is necessary to characterize the uncertainty of the random variables defining cross-section geometries.

Fabrication tolerances can be found in different international standards, including EN 1090-2 [26], AISC 303-16 [27] and AS/NZS 1365 [28], which define maximum deviations allowed from the nominal variables and provide information on the upper boundary of the investigated variabilities. Several research works studying geometric uncertainties on steel structures can be found in the literature, but they mainly focus on hot-rolled [29,30] or cold-formed [15,31] carbon steel profiles. Studies of stainless steel specimens are scarcer and generally limited to the work conducted by Afshan et al. [17] and Lin et al. [18,19] on a relatively small number of specimens. This paper investigates the variability of cross-section dimensions for a wide range of stainless steel product types available based on an extensive database collected from the literature.

The assembled database includes 1070 cold-formed Square and Rectangular Hollow Sections (SHS and RHS) [32-71], 165 cold-formed Circular Hollow Sections (CHS) [39,54,72-81], 204 welded I-sections [55,59,67,74,78,82-87], 27 welded channel sections [88-89], 260 cold-formed channel and lipped channel sections [90-98], 49 hot-rolled angle sections [99-101] and 20 cold-formed angle sections [89,102]. For all section types, the most common austenitic, ferritic, duplex and lean duplex grades were included (EN 1.4301, 1.4306, 1.4318, 1.4435, 1.4541, 1.4571, 1.4404, 1.4003, 1.4509, 1.4512, 1.4462, 1.4162, 1.4062). The random variables investigated for each cross-section type are defined in Figure 1.

Results are reported in Table 1 for the different stainless steel cross-section types, including the number of specimens used in the determination of the statistics of each random variable, the calculated mean ($\mu_{X,weighted}$), standard deviation ($\sigma_{X,pooled}$) and coefficient of variation (COV), and the most suitable statistical distribution fitting the data. The analysis indicated that in general, outer cross-section dimensions (including height, width and diameter) tend to be very close to the nominal values, while measured thicknesses differ more from the nominal values, being slightly lower on average and showing considerably higher variability, in line with the results reported in previous research works for carbon steel sections [7,23]. The statistical distributions adopted in the literature for dimensional variables include both normal [5,7] and log-normal [5] distributions. Following the methodology described in Section 2, the different databases were analysed to choose and fit the most adequate probabilistic distribution for the different random variables. Generally, the skewness values calculated for the different geometric variables indicated considerably symmetric data populations (skewness values ranged between -1 and 1), although some of the kurtosis values were found to be notably high for the outer dimensions. The comparison of the empirical and the hypothesized cumulative distribution functions and the inspection of the histograms indicated that normal distributions can be adopted for cross-section dimension random variables, as recommended in [5,7]. Typical histograms corresponding to the investigated cross-section dimensions are shown in Figure 2, representing measured-to-nominal height ratios for SHS/RHS and I-sections, and measured-to-nominal thickness ratios for CHS. In general, these results are in line with the variability measurements reported in [17-19] and in prEN 1993-1-1, Annex E [1] for equivalent stainless steel and carbon steel cross-sections, but show lower variations than the COV value of 0.05 given in [18,19].

Although the corner radius is an important parameter defining square and rectangular hollow sections, nominal values are usually not provided by producers. Recommendations can be found in the literature and in the recent Design Manual for Stainless Steel Structures [103], which indicate that the internal corner radius $r_i$ should be taken as $r_i = 2t$ if no measurements are available, where t is the section thickness. Since the collected database on SHS and RHS included measurements of the external R or the internal $r_i$ corner radius for a large number of specimens, the recommendation of using $r_i = 2t$ is assessed in this paper. It is important to note that when the corner regions of SHS and RHS specimens are examined, the thickness in these areas is not completely uniform due to the high plastic strains occurring during the forming process. Thus, assuming the external radius is equal to the internal radius plus the thickness can be inaccurate. Consequently, corner radius measurements are analysed separately for the reported external and internal radii, by normalizing data with the nominal thickness and calculating the mean and standard deviations of the measured radius-to-nominal thickness ratios, $R/t_n$ and $r_i/t_n$. The results are reported in Table 1 and indicate very similar dimensions for the corner radius regardless of whether the reference dimension is taken as the external corner radius ($R = 2t_n$) or the internal radius ($r_i = t_n$) for cold-formed SHS and RHS. It is also worth noting that the scatter observed for radii measurements is high, especially for the internal radius, owing to the difficulties associated with taking these measurements.

The correlation matrices for the cross-section dimensions are presented in Eqs. 3-7 for the different cross-section shapes investigated in this paper. While Eq. 3 presents the correlation matrix for the cross-section dimensions in SHS and RHS sections, Eq. 4-Eq. 7 report the correlation coefficients for CHS, I-sections, channel and angle sections. The results indicate that positive correlations exist between the external dimensions H and B for SHS/RHS, welded I-sections and channel sections (with correlation coefficients being between 0.20 and 0.50) and between the web and flange plate thicknesses for welded I-sections, while negative correlations are observed between the thickness and the external dimensions H and B for channel sections. Nevertheless, all correlation coefficients are found to be below 0.50.

$$\text{CORR}_{\text{SHS/RHS}} = \begin{matrix} \\ H \\ B \\ t \\ A \\ W_{pl} \end{matrix} \begin{matrix} H & B & t & A & W_{pl} \end{matrix} \begin{bmatrix} 1.000 & 0.490 & 0.187 & 0.278 & 0.393 \\ 0.490 & 1.000 & 0.072 & 0.159 & 0.238 \\ 0.187 & 0.072 & 1.000 & 0.994 & 0.973 \\ 0.278 & 0.159 & 0.994 & 1.000 & 0.991 \\ 0.393 & 0.238 & 0.973 & 0.991 & 1.000 \end{bmatrix} \quad \text{Eq. 3}$$

$$\text{CORR}_{\text{CHS}} = \begin{matrix} \\ D \\ t \\ A \\ W_{pl} \end{matrix} \begin{matrix} D & t & A & W_{pl} \end{matrix} \begin{bmatrix} 1.000 & 0.004 & 0.092 & 0.181 \\ 0.004 & 1.000 & 0.996 & 0.984 \\ 0.092 & 0.996 & 1.000 & 0.996 \\ 0.181 & 0.984 & 0.996 & 1.000 \end{bmatrix} \quad \text{Eq. 4}$$

$$\text{CORR}_{\text{I section}} = \begin{matrix} \\ H \\ B \\ t_w \\ t_f \\ A \\ W_{pl} \end{matrix} \begin{matrix} H & B & t_w & t_f & A & W_{pl} \end{matrix} \begin{bmatrix} 1.000 & 0.254 & 0.145 & 0.048 & 0.221 & 0.394 \\ 0.254 & 1.000 & -0.157 & -0.025 & 0.065 & 0.167 \\ 0.145 & -0.157 & 1.000 & 0.346 & 0.677 & 0.534 \\ 0.048 & -0.025 & 0.346 & 1.000 & 0.896 & 0.901 \\ 0.221 & 0.065 & 0.677 & 0.896 & 1.000 & 0.968 \\ 0.394 & 0.167 & 0.534 & 0.901 & 0.968 & 1.000 \end{bmatrix} \quad \text{Eq. 5}$$

$$\text{CORR}_{\text{channels}} = \begin{matrix} \\ H \\ B \\ t \\ A \\ W_{pl} \end{matrix} \begin{matrix} H & B & t & A & W_{pl} \end{matrix} \begin{bmatrix} 1.000 & 0.215 & -0.483 & -0.399 & -0.262 \\ 0.215 & 1.000 & -0.248 & -0.204 & -0.166 \\ -0.483 & -0.248 & 1.000 & 0.995 & 0.970 \\ -0.399 & -0.204 & 0.995 & 1.000 & 0.989 \\ -0.262 & -0.166 & 0.970 & 0.989 & 1.000 \end{bmatrix} \quad \text{Eq. 6}$$

$$\text{CORR}_{\text{angles}} = \begin{matrix} \\ L \\ t \\ A \end{matrix} \begin{matrix} L & t & A \end{matrix} \begin{bmatrix} 1.000 & -0.080 & 0.131 \\ -0.080 & 1.000 & 0.977 \\ 0.131 & 0.977 & 1.000 \end{bmatrix} \quad \text{Eq. 7}$$

The variability of the individual random variables defining cross-section dimensions investigated in Table 1 is fundamental for the definition of the numerical simulations for the subsequent structural reliability analyses for direct analysis methods. However, in terms of the statistical validation of proposed design approaches following the traditional *two-stage* member-based approach, the effect of individual parameters on the final geometric variation parameter $V_{geom}$ depends on the considered resistance functions $g_{rt}$ [17]. For the calculation of the geometric coefficients of variation for compression and bending, Eq. 8 and the procedure described in [17] were followed, in which $w_i$ is the weighting factor of the random variable $X_i$, calculated from Eq. 9, while $V_i$ and $\mu_i$ correspond to the coefficient of variation and mean value of the random variable $X_i$, respectively.

$$V_{geom}^2 = \sum_{i=1}^{n}(w_i V_i)^2 \qquad \text{Eq. 8}$$

$$w_i = \frac{\frac{\partial g_{rt}}{\partial X_i}\mu_i}{g_{rt}(\underline{X}_m)} \qquad \text{Eq. 9}$$

Results for the geometric variability are reported in Table 2, which show a major influence of the variability of the thickness on the calculated geometric variation parameters for both compression and bending. This follows from the fact that the calculated $V_{geom}$ parameters are very similar to the corresponding thickness COV values while the variability of outer cross-section dimension parameters is considerably lower (see Table 1). This dependence of the geometric variation parameters A and $W_{pl}$ with respect to the section thickness is also evident from the correlation matrices given in Eqs. 3-7, which show similar correlation values to the results reported in [11]. Results reported in Table 1 and Table 2 are also in line with the results reported in [17], although the estimated variabilities of the different variables are lower in this analysis since the collected database was considerably more extensive. Calculated $V_{geom}$ values are also comparable to the values reported in the literature for carbon steel sections [29], reporting $V_A = 0.02$ and $V_W = 0.02$ values for I-sections in compression and bending, respectively. Note that considerably higher values of geometric variability $V_{geom} = 0.05$ were adopted in the development of the AISC stainless steel design guide [104] and in [18].

## 4. MATERIAL PROPERTIES

### 4.1. GENERAL

The second main group of random variables governing the strength of structures are material properties. For stainless steels, material parameters are expected to introduce a higher level of uncertainty in structural reliability analysis due to the pronounced nonlinearity of the stress-strain curve, the significantly large number of alloys available, anisotropy and asymmetry of cross-sections. The material model most commonly used for the description of the nonlinear stress-strain response of stainless steel alloys is the two-stage model developed by Mirambell and Real [105] shown in Eq. 10, which was based on the Ramberg-Osgood model [106,107]. Basic nominal material properties are defined in structural [2,103] and material standards (EN 10088-1 [108] and EN 10088-4 [109]) for different stainless steel grades.

$$\varepsilon = \begin{cases} \dfrac{\sigma}{E} + 0.002 \left(\dfrac{\sigma}{f_y}\right)^n & \text{for } \sigma \leq f_y \\ \dfrac{\sigma - f_y}{E_{0.2}} + \varepsilon_{0.2} + \left(\varepsilon_u - \varepsilon_{0.2} - \dfrac{f_u - f_y}{E_{0.2}}\right)\left(\dfrac{\sigma - f_y}{f_u - f_y}\right)^m & \text{for } \sigma > f_y \end{cases} \qquad \text{Eq. 10}$$

In Eq. 10, E is the Young's modulus, $f_y$ is the yield stress (usually adopted as the 0.2% proof stress), $E_{0.2}$ is the tangent modulus at the 0.2% proof stress, $\varepsilon_{0.2}$ is the total strain at the 0.2% proof stress, $f_u$ and $\varepsilon_u$ are the ultimate strength and corresponding total strain, respectively, and *n* and *m* are the strain hardening exponents. A comprehensive analysis on material modelling of stainless steel alloys was conducted and expressions for predicting the different parameters involved in Eq. 10 were proposed in [110], and standardised representative values for the key material parameters were proposed in [111] for numerical parametric studies, although the variability with respect to their nominal values was not investigated in these research works. Therefore, it is necessary to statistically define these material parameters to allow their random variations to be modelled in the advanced analysis of stainless steel members and systems. The sensitivity of the system strength to the individual material parameters will be investigated in the future for different types of stainless steel structural systems and failure modes (i.e. yielding and instability modes).

The variability of the material properties of carbon steel structures has been extensively investigated [6-9,29-31], but only a few studies on the variability of the yield stress $f_y$ and the ultimate tensile strength $f_u$ are available for stainless steels [17,18,112]. Table 3 summarizes the overstrength ratios (i.e. measured-to-nominal yield stress and ultimate tensile strength ratios) and respective COV values recommended in the literature over the last decades for stainless steels. The latter are based on extensive databases collected from international producers and additional tests on stainless steel plate and sheet materials. Note that while the overstrength values recommended in [17,112] correspond to the nominal material properties defined in the European EN 10088-4 [109] Standard, values in [18] refer to the equivalent US ASTM A666 [113] standard. However, these studies did not incorporate other fundamental material parameters such as the Young's modulus, strain hardening parameters and the ultimate strain. With the aim of providing information on the variability of these parameters, and with no access to data from producers, an extensive database was collected from the literature including information on austenitic [36,39-42,46,52,54,55,59-62,64,67,68,71,73-76,77,80-89,96-98,100,102, 110,114-128], ferritic [32,35,37,46-48,50,51,65,66,73,90,93,96-98,110,114,115,122,129-133] and duplex/lean duplex stainless steel grades [38,43,49,57,58,62,63,65,71,73-75,78,97,110,114,122,134-138]. The database contained data on more than 1400 specimens, covering samples tested on base material (plates and sheets) and cold-formed material (flat, corner regions of SHS and RHS, and CHS). The investigated base material samples included plate thickness values ranging between 1 mm and 20 mm, while the thickness of the samples extracted from cold-formed specimens varied between 1 mm and 10 mm, with overall cross-section dimensions between 20 mm and 315 mm. This information was used in this paper to define statistical parameters and distributions for the Young's modulus, yield stress, ultimate strength, ultimate strain and strain hardening parameters, which are presented in Sections 4.2-4.6.

## 4.2. YOUNG'S MODULUS

Young's modulus or modulus of elasticity E is the slope of the initial part of the stress-strain curve, and a key parameter defining the deformation of stainless steel members and the stiffness of stainless steel structures as well as their strength when failure is governed by instability modes. For typical hot-rolled carbon steel, the slope of the stress-strain curve is constant until the yield point, but for cold-formed carbon steels and stainless steels the slope gradually decreases after the proportionality stress is reached and yielding commences. Current structural and material standards for stainless steel [2,103,108] provide slightly different nominal values for the Young's modulus: while [2,103] adopt a unique value of 200 GPa for the three main stainless steel types, [108] provides a value of 200 GPa for most common austenitic, duplex and lean duplex grades and 220 GPa for ferritic grades. Alternatively, the AS/NZS 4673 [139] and SEI/ASCE-8 [140] standards adopt the slightly lower nominal Young's modulus values of 195 GPa and 193 GPa, respectively, for the most common austenitic grades.

However, the values of the Young's modulus reported in the literature show considerably different values, as shown in Table 4, where the weighted mean, pooled standard deviations and coefficients of variation of the Young's modulus are provided separately for plates and sheets and for cold-formed material. The results indicate that the nominal values provided in European standards [2,103,108] tend to slightly overestimate the Young's modulus, especially for austenitic grades, although measurements seem to be more aligned with the nominal values provided in Australian and US design codes [139,140]. According to the different publications gathered, approximately half of the material test programmes were conducted in accordance with the ISO 6892-1 [141] standard, while the remaining authors followed the recommendations in the AS 1391 [142] and ASTM E8M [143] standards, which are the reference standards according to which the Young's modulus values in EN 10088-1 [108], AS/NZS 4673 [139] and SEI/ASCE-8 [140] are determined, respectively. However, while the test rates adopted in product certifications tend to be on the upper limits of the test rate ranges provided in the standards, the rates used by researchers in laboratory tests are generally towards the lower bound, which are considered to be more similar to the loading rates in real structures. While the lower values of Young's modulus appear to be correlated with slower test rates, it is generally accepted that the Young's modulus is strain rate independent. From Table 4 it can also be appreciated that cold-forming processes have little effect on the value of the Young's modulus, as anticipated in [144], showing similar levels of variability to those observed for plates and for the entire database. Although statistical studies including results on the Young's modulus are scarce in the literature, research works by Hess et al. [6] suggested that a normal distribution should be adopted to model the variability of the elastic modulus. Data on Young's modulus was found to be considerable symmetric for stainless steel alloys (with skewness values ranging between -1.15 and -0.69), as illustrated by the Young's modulus histograms shown in Figure 3 for different stainless steel types including plate and cold-formed material, which suggest that the adoption of normal distributions is reasonable for the three material types.

## 4.3. YIELD STRESS $f_y$

Characterized by gradual yielding, stainless steels do not present a clearly defined yield stress as for common structural steel, and thus the 0.2% proof stress is conventionally adopted as the equivalent yield stress $f_y$. Among the different mechanical characteristics of steels, the yield strength is the parameter most pervasively used in design. Minimum specified yield stress values are defined in the corresponding material Standard EN 10880-4 [109], defined as the characteristic values corresponding to the 95% confidence limit. The same characteristic values are assumed to represent the nominal values according to EN 1990 [145]. Table 5 reports the statistical data for the variability of the yield stress values collated in this paper for plate and sheet specimens. Although the number of data represented a smaller proportion of the vast databases analysed in [17] and [112], calculated overstrength ratios and COV values show a good agreement with the values reported in Table 3, especially for the most recent values recommended in [112]. The data analysed also suggest that a log-normal function can be adopted to describe the distribution of the yield stress, which is in line with the recommendations provided by the JCSS [5]. The overstrength ratios provided in [17] have been utilized in the reliability calibrations of several design provisions for the European EN 1993-1-4 [2] Standard, including expressions for both plated structures and cold-formed specimens.

Table 6 shows the variability of the yield stress for cold-formed materials, including the overstrength ratios resulting from using the reference nominal yield stress values provided in EN 10880-4 [109]. Results in Table 6 demonstrate that the measured-to-nominal ratios for cold-formed materials are considerably higher and more scattered than for plate and sheet specimens (see Table 5), which can be attributed to the strength enhancements arising from the plastic strains occurring during the cold-forming process. In order to achieve more consistent overstrength ratios for all stainless steel product types, the measured yield stresses were also normalized by the enhanced nominal yield stresses calculated from the predictive enhancement models proposed in [103,144] for cold-formed specimens, which are based on the nominal material and cross-section properties. The results are also included in Table 6, in which lower overstrength ratios and coefficients of variation can be observed. Although it is considered that the overstrength ratios based on extensive databases for plate and sheet material properties can be conservatively adopted for cold-formed materials, by assuming that the higher scatter will be compensated by the considerably higher overstrength ratios, the recommendation, when a more accurate characterization of the variability of the yield stress is required, is to adopt the reference enhanced yield stress based on nominal properties in conjunction with overstrength ratios and COV values equal to 1.30, 1.30, 1.15 and 0.15, 0.15, 0.15 for cold-formed austenitic, ferritic and duplex stainless steels, respectively. It is important to highlight that the overstrength ratios reported in this paper (see Table 5 and Table 6) correspond to the EN 10880-4 [109] minimum specified yield stress values. Since the ASTM A666 [113] minimum specified values for equivalent stainless steel grades are somehow lower, equivalent overstrength ratios would be around 10% higher if based on the ASTM A666 [113] minimum specified values.

Figure 4 shows the histograms for the overstrength ratios for the yield stress for the cold-formed stainless steel families, referred to as the enhanced nominal yield stress values. The assembled data showed a slight positive skewness since the lower tails of the distributions representing the yield stress are typically affected by the rejection of materials which do not meet the minimum specified values required. For this reason, log-normal distributions are usually suggested to fit yield stress data in the literature [5,6,8,9]. Following a similar approach, the data collected in this study has been fitted by log-normal distributions, which were found to be adequate by the Anderson-Darling tests and the histograms shown in Figure 4. Results for plate and sheet materials also showed that the most suitable statistical distribution was a log-normal distribution.

### 4.4. ULTIMATE TENSILE STRENGTH $f_u$

The ultimate tensile strength $f_u$ corresponds to the maximum stress value of the stress-strain curve prior to failure. Although less significant than the yield stress because it is generally not a design variable for structural members, it provides an indirect measurement of the level of strain hardening characterizing stainless steels. Following the same procedure as for the yield stress, measured ultimate strength $f_u$ values from the assembled database were compared with the corresponding nominal values given in EN 10088-4 [109] for each grade. As for yield stress, the nominal (minimum specified) ultimate strength $f_{u,nom}$ is the characteristic value corresponding to the 95% confidence limit. Statistical data on the variability of the ultimate strength is presented in Table 7 for the different stainless steel types investigated in this paper, with results for the different product types reported separately. Note again that the overstrength ratios reported correspond to the EN 10880-4 [109] minimum specified tensile strength. Equivalent overstrength ratios based on ASTM A666 [113] minimum tensile strengths would be higher than the values reported in Table 7 since the minimum specified tensile strength values are lower, in general, for ASTM A666 than for EN 10880-4. According to the results published in [17], the overstrength factors for the ultimate strength $f_{u,mean}/f_{u,nom}$ ranged between 1.06 and 1.23 for plate and sheet materials, with no clear correlation with the associated COV (referred to the EN 10880-4 [109] minimum specified tensile strengths). Although the databases utilized for the recommendations in [17,112] were considerably more extensive than the sample population analysed in this paper, results show that the calculated overstrength ratios for plate and sheet materials are in line with those reported in Table 3. Results also suggest that cold-formed materials present considerably more scattered overstrength ratios than plates and sheets, showing higher overstrength ratios for austenitic stainless steel alloys, but similar values for ferritic and duplex grades. From these results, it is recommended that the same overstrength factor provided for plates and sheets also be adopted for cold-formed materials, and the associated COV values be increased to 0.10 to account for the more scattered results. In line with the findings reported in the previous section for the yield stress, results for the ultimate tensile strength also tend to show a positive skewness due to the rejection of materials that do not comply with the minimum specified values. Consequently, log-normal distributions are generally adopted in the literature to model the ultimate tensile strength [5,6,8]. Thus, and following the methodology described in Section 2, the data was accurately fitted by log-normal distributions in all cases, as shown in Figure 5 for cold-formed stainless steel materials.

## 4.5. ULTIMATE STRAIN $\varepsilon_u$

The ultimate strain $\varepsilon_u$ is one of the material parameters most affected by the plastic deformations occurring during forming processes, decreasing considerably when compared to the corresponding base sheet or plate material values [110]. Statistical parameters of the ultimate strain for plates and sheets and cold-formed material are reported separately in Table 8 for austenitic, ferritic and duplex stainless steel alloys. These results are equivalent to those proposed in [111] for plates and sheets and slightly higher for cold-formed materials, and highlight the reduction observed in the ultimate strain for cold-formed coupons and indicate that austenitic stainless steels show the highest deformation capacity, followed by duplex and ferritic grades. Although log-normal distributions have been suggested in the literature for the ultimate strain [5,10], the skew and kurtosis values calculated from the assembled data ranged between -0.7–1.1 and 2.0–4.6, suggesting that the data could be accurately fitted by normal distributions. Anderson-Darling normality tests confirmed the selection of these distributions.

4.6. STRAIN HARDENING PARAMETERS $n$ AND $m$

The roundness of the nonlinear stress-strain curve of stainless steel alloys is controlled by the strain hardening exponent $n$ for the low strain range, and by the strain hardening exponent $m$ for the high strain range, being key parameters in the material definition of stainless steels. These parameters cannot be directly measured from experimental stress-strain curves; rather they need to be optimized or estimated through numerical techniques. The strain hardening exponent values considered in this paper were optimized following the procedure described in [130] for those stress-strain curves acquired as raw data, whereas the values reported in the literature were otherwise assumed.

Table 9 summarizes the weighted mean, pooled standard deviations and coefficients of variation for the strain hardening parameter $n$. Results are provided separately for different stainless steel and product types, since alloying elements and cold-working tend to considerably influence the parameter [110]. Although to the authors' knowledge no studies are available in the literature on the statistical characterization of the strain hardening parameters, the literature recommends the adoption of log-normal distributions for material-related parameters [5]. The database for the strain hardening parameter $n$ showed a slightly positive skewness, suggesting that log-normal distributions would be adequate to fit the data. This can be explained by the close relationship between the strain hardening parameter $n$ and the yield stress. Thus, the data for the strain hardening parameter $n$ was fitted with log-normal distributions, as shown in the histograms presented in Figure 6 for different cold-formed stainless steel types, which were confirmed to be suitable by Anderson-Darling tests subsequently carried out. From the statistical results reported in Table 9, it can be also concluded that the cold-working process induces a slightly more rounded stress-strain response, providing lower $n$ values for cold-formed specimens than for plates and sheets. This is more relevant for ferritic grades, which show the highest $n$ values. Note that the mean values reported in Table 9 for plates and sheets are slightly higher than the nominal values provided in [2,103], while values for cold-formed material are lower than the nominal values. This is because these nominal values were based on the analysis conducted in [110], in which, for simplicity, no distinction was made between product type in the proposed $n$ values.

Similar results are presented in Table 10 for the strain hardening parameter $m$, in which the effect of the cold-forming is also evident for all stainless steel types. However, higher values were observed for cold-formed specimens in this case, due to the reduction of the ultimate strain and the slight increase in ultimate strength occurring during the cold-forming process, which cause the increase in the strain hardening parameter $m$ observed from the data. Although more scattered, results were also positively skewed and log-normal distributions were also observed to adequately fit the collected data by the Anderson-Darling tests carried out. The mean values of the strain hardening parameters obtained in this study are similar to the values recommended in [111] for parametric studies.

4.7. CORRELATIONS BETWEEN MATERIAL PARAMETERS

The correlation matrices corresponding to the material properties analysed in this paper are presented in this Section. The correlation matrix corresponding to austenitic stainless steel alloys is provided in Eq. 11, while the matrices in Eq. 12 and Eq. 13 report the correlation coefficients for ferritic and duplex stainless steels, respectively. From the calculated correlation matrices, it is evident that positive correlation exists between the yield stress and the ultimate tensile strength values, while negative correlations can be observed between the yield stress and the ultimate strain, and between the ultimate tensile strength and the ultimate strain for all stainless steel families, in line with the correlations reported in the literature for structural steels [5,11]. In addition, positive correlations between the strain hardening parameter *m* and the yield stress and the ultimate tensile strength are highlighted, as are negative correlations between the strain hardening parameters *n* and *m*.

$$\text{CORR}_{\text{austenitic}} = \begin{matrix} & \begin{matrix} E & f_y & n & f_u & \varepsilon_u & m \end{matrix} \\ \begin{matrix} E \\ f_y \\ n \\ f_u \\ \varepsilon_u \\ m \end{matrix} & \begin{bmatrix} 1.000 & -0.122 & 0.048 & -0.124 & 0.271 & -0.160 \\ -0.122 & 1.000 & -0.341 & 0.790 & -0.757 & 0.463 \\ 0.048 & -0.341 & 1.000 & -0.341 & 0.143 & -0.300 \\ -0.124 & 0.790 & -0.341 & 1.000 & -0.527 & 0.358 \\ 0.271 & -0.757 & 0.143 & -0.527 & 1.000 & -0.457 \\ -0.160 & 0.463 & -0.300 & 0.358 & -0.457 & 1.000 \end{bmatrix} \end{matrix} \quad \text{Eq. 11}$$

$$\text{CORR}_{\text{ferritic}} = \begin{matrix} & \begin{matrix} E & f_y & n & f_u & \varepsilon_u & m \end{matrix} \\ \begin{matrix} E \\ f_y \\ n \\ f_u \\ \varepsilon_u \\ m \end{matrix} & \begin{bmatrix} 1.000 & -0.063 & -0.056 & -0.082 & 0.113 & -0.223 \\ -0.063 & 1.000 & -0.366 & 0.733 & -0.720 & 0.405 \\ -0.056 & -0.366 & 1.000 & -0.333 & 0.490 & -0.570 \\ -0.082 & 0.733 & -0.333 & 1.000 & -0.503 & 0.385 \\ 0.113 & -0.720 & 0.490 & -0.503 & 1.000 & -0.622 \\ -0.223 & 0.405 & -0.570 & 0.385 & -0.622 & 1.000 \end{bmatrix} \end{matrix} \quad \text{Eq. 12}$$

$$\text{CORR}_{\text{duplex}} = \begin{matrix} & \begin{matrix} E & f_y & n & f_u & \varepsilon_u & m \end{matrix} \\ \begin{matrix} E \\ f_y \\ n \\ f_u \\ \varepsilon_u \\ m \end{matrix} & \begin{bmatrix} 1.000 & 0.143 & -0.073 & 0.277 & 0.058 & -0.107 \\ 0.143 & 1.000 & -0.045 & 0.855 & -0.624 & 0.263 \\ -0.073 & -0.045 & 1.000 & -0.277 & 0.181 & -0.283 \\ 0.277 & 0.855 & -0.277 & 1.000 & -0.534 & 0.094 \\ 0.058 & -0.624 & 0.181 & -0.534 & 1.000 & -0.402 \\ -0.107 & 0.263 & -0.283 & 0.094 & -0.402 & 1.000 \end{bmatrix} \end{matrix} \quad \text{Eq. 13}$$

## 5. INITIAL GEOMETRIC IMPERFECTIONS

Random variables related to initial geometric imperfections represent the third fundamental source of uncertainty to be considered in the analysis and design of stainless steel structures. Their effect is to introduce additional load eccentricities and increase second order effects, causing the premature failure of the structure. Initial geometric imperfections inevitably occur in stainless steel structures despite the strict manufacturing and installation tolerance requirements, and thus imperfections need to be explicitly introduced in finite element models, or implicitly incorporated through the analytical resistance functions provided in national codes for the current *two-step* member-based design approach. For the calibration of the direct design methods assisted by finite element analysis, imperfections need to be included and treated as random variables to account for these uncertainties in the reliability studies. Statistics corresponding to local cross-sectional, member and global frame imperfections are analysed in the following sub-sections for stainless steel structures.

## 5.1. LOCAL IMPERFECTIONS

Local cross-section imperfections are defined as deviations of the different plate elements composing the cross-section from the perfect geometry, and include distortional-shaped and local-shaped modes. For cold-formed hollow sections (SHS, RHS and CHS) the latter are more relevant since the transverse stiffness of hollow sections is considerably high, whereas for open sections (including I-sections and channels) both types can be important.

The amplitude of local imperfections $w_{0,local}$ in carbon steel sections has traditionally been estimated by means of different predictive models, the simplest of which define the local imperfection amplitude as a fraction of the wall thickness [5,15], while deviation tolerances given in EN 1090-2 [26] are expressed as fractions $\gamma_1$ of plate width b ($\pm b/\gamma_1$), with $\gamma_1 = 100$ for welded sections and $\gamma_1 = 50$ for cold-formed sections. However, the adoption of the local imperfection factor as a fraction of the plate thickness or width was found to be unsuitable as a general parameter for cold-formed sections, and the alternative model given by Eq. 14 was developed by Dawson and Walker [146] for simply supported plates in carbon steel, with $\gamma_2 = 0.2$. In this equation $f_y$ is the yield stress (or the 0.2% proof stress) and $\sigma_{cr}$ corresponds to the plate critical buckling stress. This model was adapted to stainless steel SHS and RHS sections by Gardner and Nethercot [147], where the coefficient $\gamma_2 = 0.023$ was proposed. Despite being initially based upon just 15 imperfection measurements, it has subsequently been found to be adequate by different researchers [36,57,65].

$$\frac{w_{0,local}}{t} = \gamma_2 \frac{f_y}{\sigma_{cr}} \qquad \text{Eq. 14}$$

Since no statistical study is available in the literature for local imperfections on stainless steel cross-sections, despite the considerable database of reported local imperfection measurements available in the literature, this paper presents a comprehensive analysis of local imperfection amplitudes for different cross-section shapes. The collected database includes measurements for cold-formed SHS and RHS sections [32-37,42,43,47,49,57,58,62,64-66,147], welded I-sections [83-86], welded channel sections [88-89], cold-formed channel and lipped channel sections [97], hot-rolled angle sections [99-101] and cold formed angle sections [102], including specimens made from different stainless steel families. The analysis is presented separately for welded and cold-formed sections. Results corresponding to the calculated $\gamma_1$ factors for welded I-sections, channels and angles are reported in Table 11, where the different geometric properties $B_n$ and $H_n$ correspond to those defined in Figure 1 for each section type. The skewness and kurtosis values calculated for the $\gamma_1$ factors suggested that they follow a normal distribution for I-sections and a log-normal distribution for angle sections. This was further confirmed by Anderson-Darling tests. On the contrary, the data available for channel sections did not allow determining an appropriate statistical distribution and thus only the mean and standard deviation values are reported in this paper.

The data on local imperfection amplitudes for cold-formed stainless steel sections is plotted in Figure 7, where it is compared with the original predictive model (Eq. 14 with $\gamma_2 = 0.023$). The results show that data on SHS and RHS sections can be grouped in two sets, named Group 1 and Group 2, showing considerably different imperfection amplitudes. While data for Group 1 is similar to the imperfection measurements reported for angle and channel sections, lying close to the original proposal (Eq. 14 with $\gamma_2 = 0.023$), Group 2 exhibits considerably larger imperfections. This difference can be explained by specimens being supplied by different stainless steel producers with different quality control requirements, and by the adoption of different procedures for the measurement of imperfection amplitudes. A review of the data sources indicated that most specimens were delivered by the same producer, but results for each data Group corresponded to different measurement techniques: while results included in Group 1 [32-37,42,57,58,64-66,147] were acquired from longitudinal measurements at the different faces comprising SHS and RHS members, measurements in Group 2 [43,47,49,62] were obtained through perimetral (transversal) readings at different cross-sections along the member. In this paper, the results are treated independently for the two Groups. Regression fits for SHS/RHS Group 2 and data for other sections can be observed in Figure 7. It can be noticed that while the regression fit for Group 1, with $\gamma_2 = 0.024$, is very similar to the original proposal [147], the factor for Group 2 is considerably higher, $\gamma_2 = 0.094$. In view of these results, and for simplicity, it is recommended that the original factor $\gamma_2 = 0.023$ is adopted for the prediction of local imperfection amplitudes of cold-formed sections when no measurement is available in deterministic analyses. Statistical information on the $\gamma_2$ parameter for the different cold-formed cross-sections is provided in Table 12. For all groups, the $\gamma_2$ parameters were found to be positively skewed and so log-normal distributions were fitted, and the goodness-of-fit of these distributions was confirmed by the corresponding Anderson-Darling tests.

## 5.2. MEMBER IMPERFECTIONS

Member imperfections are bow-shaped deviations from the perfect geometry in which cross-sections are displaced or rotate as whole, and are generally associated with global buckling shapes. These initial member imperfections have a considerable effect on member capacity as they increase second order effects and can cause premature failure. They are implicitly considered in the resistance functions provided in current standards [1-4], such as through the adoption of buckling curves for the flexural buckling resistance of columns. However, for the reliability studies required for the derivation of system safety factors these inevitable imperfections need to be explicitly characterized and included in the finite element simulations.

Member imperfections have been extensively investigated for hot-rolled and cold-formed steel members [5,15,29] and are traditionally represented by a half sine-wave with a maximum amplitude given by a fraction K of the member length L, $w_{0,member} = L/K$. Despite the considerable database available from international research groups reporting measured imperfection values from experimental specimens during the last decades, to the authors' knowledge no equivalent studies can be found in the literature for stainless steel columns. Thus, reported bow imperfection amplitudes at the member mid-length section, where the maximum deviation is most likely to appear, were collected from the literature on austenitic, ferritic, duplex and lean duplex stainless steel members with different cross-section types, and the analysis is presented in this section. Weighted mean, pooled standard deviations and coefficients of variation corresponding to the mid-length member imperfection amplitude are reported in Table 13 in terms of the length fraction K, while histograms for SHS/RHS, CHS and I-sections are shown in Figure 8. According to the results reported in Table 13, the measured member imperfections tend to be considerably lower than the corresponding tolerances stipulated in EN 1090-2 [26] (i.e. L/750) and imperfection values adopted in the development of the buckling curves in European Standards [1,2] (i.e. L/1000). Values are also considerably lower on average than those regularly adopted in numerical simulations [42,57,66,73] (i.e. L/1000, L/1500), but it should be noted that initial geometric imperfections are sometimes used to indirectly incorporate the combined effect of different types of imperfections in numerical simulations, such as residual stresses. Even though the normal distribution has been postulated as more acceptable for initial geometric imperfections in members by some research works [5,148], the log-normal distribution has also been found to be adequate [149]. The positively skewed data (with skewness values ranging between 1.5 and 2.4), with kurtosis values above 3.0 suggests that it can be reasonably fitted by log-normal distributions for the different cross-section types investigated. This selection was confirmed by Anderson-Darling tests and the histograms shown in Figure 8.

However, assuming member imperfections as half sine-waves (represented by the first buckling mode) in conjunction with the mid-length amplitude value does not provide information about the contribution of higher buckling modes showing multiple waves [148]. Following the procedure described in [148], initial member geometric imperfections were assumed to be a linear combination of a given number of modes *n*, with individual buckling modes being sine-shaped with different amplitude factors as per Eq. 15. In this equation $a_i$ is the scale factor for the $i^{th}$ mode and x is the non-dimensional coordinate along the length of the member.

$$w_{0,member} = \sum_{i=1}^{n} L \cdot a_i \sin(i\pi x) \qquad \text{Eq. 15}$$

To treat member imperfections as random variables defined by random shapes and magnitudes, detailed measurements along the member are required, which are not commonly measured nor reported in the literature. Thus, the database on continuous measurements of member imperfections assembled in this paper is more limited, including data on SHS and RHS [35,36,120], CHS [78] and lipped channel [96] sections. Based on these data, initial member imperfections were modelled as linear combinations of the first three buckling modes (*n* = 3 in Eq. 15), from which scale factors were determined by solving a system of equations for each member. Mean scale factors were then determined to match the mean imperfection amplitudes reported in Table 13, which represented a larger sample of mid-length measurements, and taking into account that only the first and the third modes contribute to the imperfection in this section, as described in [148]. Table 14 presents the statistical information about the obtained scale factors for the different cross-section types investigated. According to the results, the first buckling mode governs the member imperfections of the cross-section shapes investigated and should be sufficient for an accurate approximation of initial imperfections, with $a_1$ amplitudes equivalent to L/3333, L/3845 and L/5882 for SHS/RHS, CHS and channel sections, respectively. Scale factors for the first buckling modes $a_1$ were found to follow log-normal distributions, similarly to the results shown in Table 13 and Figure 8, while the distributions of $a_2$ and $a_3$ were approximately normal [148].

## 5.3. FRAME IMPERFECTIONS

Frame imperfections are deviations of the overall structure from the perfect geometry, the most representative of which are sway-shaped imperfections. These sway imperfections are usually defined by an out-of-plumb angle and have considerable influence on the strength of structures as they increase second order effects. Maximum column inclination deviations of portal frames according to the erection tolerances given in EN 1090-2 [26] are $\phi = \pm 1/500$, although designers often define more strict tolerances for serviceability and operational reasons. Owing to the importance of these imperfections, they need to be incorporated in frame analyses as initial imperfections or equivalent horizontal loads, as per prEN 1993-1-1 [1], with the out-of-plumb angle calculated from a basic angle of $\phi_0 = \pm 1/200$ and correction factors for height and number of columns in a row.

Contrary to the other imperfection types investigated in this paper, measurements on stainless steel frame imperfections are very scarce. Arrayago et al. [36] recently conducted an experimental programme on stainless steel frames, in which four single-bay single-storey frames made from RHS austenitic stainless steel were tested under static vertical and horizontal loads. Initial imperfections were obtained prior to testing by means of a precision theodolite, measuring the position of several points in each column and along the rafters. In-plane imperfections as well as out-of-plane imperfections were measured and the statistics of the recorded out-of-plumb angles are reported in Table 15. According to [36], two of the frames showed imperfection patterns following sway modes, while the initial geometries of the remaining two frames showed non-sway modes. Note that no statistical distribution is provided in this case, since the data available was too limited.

To characterize frame imperfection uncertainties adequately it is thus necessary to refer to the measurements conducted on carbon steel frames, as discussed in [15], which are also considerably limited. Based on the measurements of initial frame imperfections on general steel structures found in [150,151], a normal out-of-plumb angle distribution with a zero mean and a standard deviation of 1/610 was proposed. It is believed that the distributions provided for frame imperfection amplitudes in cold-formed carbon steel represent an upper bound for stainless steel frames, since workshops for stainless steel structures are generally more specialized and therefore final structures will typically present imperfections with lower amplitude and level of uncertainty.

## 6. RESIDUAL STRESSES

Residual stresses are stresses that exist in structural sections in their unloaded state, which are primarily generated during the fabrication processes and caused by either differential cooling of different parts of the cross-section (in hot-rolled and welded sections) or non-uniform plastic deformations (in cold-formed sections). They are particularly relevant in frames prone to instability failure modes, as they can cause premature yielding and buckling, thus reducing the ultimate strength. Owing to different material and thermal properties, the residual stresses are different in equivalent stainless steel and carbon steel sections and thus need to be independently characterized. Over the last few decades a number of experimental studies have been carried out with the aim of developing residual stress models for different stainless steel section types and fabrication processes, including cold-formed SHS and RHS sections [43,62,152-155], press-braked and hot-rolled angle sections [152,153], welded I-sections [85-87,156] and welded SHS and RHS sections [156], among others.

The residual stress pattern and magnitude vary significantly for different cross-section types, fabrication procedures and material grades, and hence different predictive models can be found in the literature for cold-formed and welded RHS, as well as for austenitic and duplex welded RHS. Consequently, the residual stress data available in the literature were sorted in different sets and only those showing a statistically representative number of specimens were included in the analysis. For the characterization of the uncertainties associated with residual stresses in stainless steel members, a similar approach to that adopted in [157] was followed. In this approach, the residual stress pattern was assumed to be deterministic for each cross-section type and fabrication process, while the magnitudes of the residual stresses were considered random variables. For the different measurements available in the literature, appropriate residual stress patterns were fitted to experimental residual stresses by applying a random scale factor. These scale factors corresponding to each specimen $j$ were calculated according to the method described in [157], which can be summarized as follows: (i) nondimensionalise measured residual stresses by the corresponding yield stress $\sigma_{exp,i}/f_y$; (ii) calculate the theoretical nondimensionalised residual stress for the considered model at the same points $\sigma_{model,i}/f_y$; (iii) calculate scale factors $Y_j$ by minimising the total error between the scaled theoretical model and the experimental measurement, as per Eq. 16.

$$\text{Error}_j = \sum_{i=1}^{n} \left( Y_j \frac{\sigma_{model,i}}{f_y} - \frac{\sigma_{exp,i}}{f_y} \right)^2 \qquad \text{Eq. 16}$$

The reference residual stress patterns considered in this study are the models proposed by Gardner and Cruise [153] for cold-formed SHS and RHS and press-braked angle sections, and the models developed by Yuan et al. [156] for welded I-sections and SHS and RHS sections, as summarized in Table 16. Note that some of the residual stress measurements from the literature review provided at the beginning of this section have not been considered in the analysis, since the number of available data was deemed insufficient for an adequate statistical characterization. The skewness and kurtosis values calculated from the residual stress scale factors ranged between -0.81 and 0.13 and 1.70 and 3.06, respectively, which together with the visual inspection of the histograms shown in Figure 9 and the Anderson-Darling tests carried out, suggest that a normal distribution is the most suitable distribution for the different stainless steel members investigated. The limited literature on the variability of residual stresses also suggests a normal distribution for the scale factors $Y_j$ [157].

# 7. CONCLUSIONS

With the purpose of providing input data for member- and system-based reliability calibrations, this paper presents a comprehensive study of the main random variables influencing the strength of stainless steel structures, including geometric properties, material parameters, imperfections and residual stresses. Based on an extensive database collected from the literature, statistical functions and probabilistic models are provided for different random variables affecting the behaviour and strength of stainless steel cold-formed and welded cross-sections (including square and rectangular hollow sections, circular hollow sections, I-sections, channels and angles) and the most common austenitic, ferritic, duplex and lean-duplex stainless steel alloys. The collection of geometric properties (section height, width, thickness and corner radius), material parameters (Young's modulus, yield stress, ultimate tensile strength, ultimate strain and strain hardening exponents), imperfections (at local cross-section, member and frame levels) and residual stresses was analysed statistically and appropriate statistical distributions were derived.

In general, height and width variations were found to be small, while variations in section thickness were significantly greater. All random variables describing geometric properties were found to be normally distributed and the analysis of the geometric data showed strong correlations between the cross-section thickness and the magnitudes governing section resistances, the cross-section area and the plastic modulus. Material parameters showed considerably higher variabilities than geometric variables, with high overstrength ratios observed for the yield stress and ultimate strength for austenitic, ferritic and duplex alloys. Considerable positive correlations were found between the yield stress and the ultimate strength for all stainless steel families, while negative correlations were obtained between the ultimate strain and the yield stress or the ultimate strength. Some material parameters were found to be normally distributed (the Young's modulus and the ultimate strain), while log-normal distributions were proposed for the remaining parameters. Finally, measured local and member imperfections were also found to follow log-normal distributions, while the data on residual stress magnitudes showed a normal variability.

The information presented in this paper will be fundamental for the development of a system-based reliability framework for stainless steel structures. However, future reliability analyses for the ultimate and serviceability limit states of stainless steel structures should also include additional uncertainties associated to the developed numerical models, connections, loads and load combinations.

**ACKNOWLEDGEMENTS**

This research project has received funding from the European Union's Horizon 2020 Research and Innovation Programme under the Marie Sklodowska-Curie Grant Agreement No. 842395. The time dedicated by numerous authors of referenced papers to provide additional data and information is also much appreciated.

**REFERENCES**

[1] European Committee for Standardization (CEN). prEN 1993-1-1. Eurocode 3: Design of Steel Structures – Part 1-1: General Rules and Rules for Buildings. Final Document; 2019.

[2] European Committee for Standardization (CEN). EN 1993-1-4:2006+A1:2015. Eurocode 3: Design of Steel Structures – Part 1-4: General Rules. Supplementary Rules for Stainless Steels, including amendment A1 (2015). Brussels, Belgium, 2015.

[3] Standards Australia (AS). AS 4100. Steel Structures. Sydney, Australia, 2016.

[4] American Institute of Steel Construction (ANSI/AISC). AISC 360-16. Specification for Structural Steel Buildings. Illinois, USA, 2016.

[5] Joint Committee on Structural Safety (JCSS). Probabilistic Model Code, Zurich, Switzerland, 2001.

[6] Hess P.E., Bruchman D., Assakkaf I.A. and Ayyub B.M. Uncertainties in material and geometric strength and load variables. *Naval Engineers Journal* 114 (2), 139–166, 2002.

[7] Kala Z., Melcher J. and Puklický L. Material and geometrical characteristics of structural steels based on statistical analysis of metallurgical products. *Journal of Civil Engineering and Management* 15, 299–307, 2009.

[8] Schmidt B.J. and Bartlett F.M. Review of resistance factor for steel: data collection. *Canadian Journal of Civil Engineering* 29, 98–102, 2003.

[9] Karmazínová M. and Melcher J. Influence of Steel Yield Strength Value on Structural Reliability. *Recent Researches in Environmental and Geological Sciences* 441–446, 2010.

[10] Ravindra M.K. and Galambos T.V. Load and resistance factor design for steel. *Journal of the Structural Division, ASCE* 104, 1978.

[11] Melcher J., Kala Z., Holický M., Fajkus M. and Rozlívka L. Design characteristics of structural steels based on statistical analysis of metallurgical products. *Journal of Constructional Steel Research* 60, 795–808, 2004.

[12] Zhang H., Shayan S., Rasmussen K.J.R. and Ellingwood B.R. System-based design of planar steel frames, I: Reliability framework. *Journal of Constructional Steel Research* 123, 135–143, 2016.

[13] Zhang H., Shayan S., Rasmussen K.J.R. and Ellingwood B.R. System-based design of planar steel frames, II: Reliability results and design recommendations. *Journal of Constructional Steel Research* 123, 154–161, 2016.

[14] Sena Cardoso F., Rasmussen K.J.R. and Zhang H. System reliability-based criteria for the design of steel storage rack frames by advanced analysis: Part I – Statistical characterisation of system strength. *Thin-Walled Structures* 141, 713–724, 2019.

[15] Sena Cardoso F., Zhang H., Rasmussen K.J.R. and Yan S. Reliability calibrations for the design of cold-formed steel portal frames by advanced analysis. *Engineering Structures* 182, 164–171, 2019.

[16] Wang C., Zhang H., Rasmussen K.J.R., Reynolds J. and Yan S. Reliability-based limit state design of support scaffolding systems. *Engineering Structures* 216, 110677, 2020.

[17] Afshan S., Francis P., Baddoo N.R. and Gardner L. Reliability analysis of structural stainless steel design provisions. *Journal of Constructional Steel Research* 114, 293–304, 2015.

[18] Lin S.-H., Yu W.-W and Galambos T.V. Load and resistance factor design of cold-formed stainless steel statistical analysis of material properties and development of the LRFD provisions. Center for Cold-Formed Steel Structures Library. 72, University of Missouri-Rolla, Missouri, U.S.A., 1988.

[19] Lin S.-H., Yu W.-W and Galambos T.V. ASCE LRFD Method for Stainless Steel Structures. *Proceedings of the 10$^{th}$ International Specialty Conference on Cold-Formed Steel Structures*. Missouri, U.S.A., 1990.

[20] MATLAB version 9.8.0 (R2020a). The MathWorks Inc., Natick, Massachusetts, 2020.

[21] Minitab version 18. Reference Manual. Minitab, LLC. State College, PA, USA, 2019.

[22] Lefebvre M. Applied Probability and Statistics. Springer Science+Business Media LLC, 2006.

[23] Melchers R.E. and Beck A.T. Structural Reliability, Analysis and Prediction. Third edition. John Wiley & Sons Ltd, 2018.

[24] Ellingwood B.R., MacGregor J.G., Galambos T.V. and Cornell C.A. Probability based load criteria: Load factors and load combinations. *Journal of the Structural Division (ASCE)* 108(5), 978–997, 1982.

[25] Ellingwood B.R. and Tekie P.B. Wind load statistics for probability-based structural design. *Journal of Structural Engineering (ASCE)* 125(4), 453–463, 1999.

[26] European Committee for Standardization (CEN). EN 1090-2:2018. Execution of steel structures and aluminium structures. Technical requirements for steel structures. Brussels, Belgium, 2018.

[27] American Institute of Steel Construction (ANSI/AISC). AISC 303-16. Code of Standard Practice for Steel Buildings and Bridges. Illinois, USA, 2016.

[28] Australian/New Zealand Standards (AS/NZS). AS/NZS 1365:1996 Tolerances for flat-rolled steel products. NSW, Australia and Wellington, New Zealand, 1996.

[29] Rang T.N., Galambos T.V. and Yu W.W. Load and resistance factor design of cold-formed steel - Statistical analysis of mechanical properties and thickness of materials combined with calibrations of

the AISI design provisions on unstiffened compression elements and connections. Structural Series, University of Missouri-Rolla, Rolla, Missouri, 1979.

[30] Byfield M.P. and Nethercot D.A. An analysis of the true bending strength of steel beams. *Proceedings of the Institution of Civil Engineers: Structures and Buildings* 128, 188–197, 1988.

[31] Foley C. Statistics on the thickness of cold-formed hollow sections. Report to AISC, 2011.

[32] Afshan S. and Gardner L. Experimental Study of Cold-Formed Ferritic Stainless Steel Hollow Sections. *Journal of Structural Engineering (ASCE)* 139, 717–728, 2013.

[33] Arrayago I. and Real E. Experimental Study on Ferritic Stainless Steel RHS and SHS Cross-sectional Resistance Under Combined Loading. *Structures* 4, 69–79, 2015.

[34] Arrayago I. and Real E. Experimental study on ferritic stainless steel simply supported and continuous beams. *Journal of Constructional Steel Research* 119, 50–62, 2016.

[35] Arrayago I., Real E. and Mirambell E. Experimental study on ferritic stainless steel RHS and SHS beam-columns. *Thin-Walled Structures* 100, 93–104, 2016.

[36] Arrayago I., González de León I., Real E. and Mirambell E. Tests on stainless steel frames. Part I: preliminary tests and experimental setup. *Thin-Walled Structures*, 2020 (in press).

[37] Bock M., Arrayago I. and Real E. Experiments on cold-formed ferritic stainless steel slender sections. *Journal of Constructional Steel Research* 109, 13–23, 2015.

[38] Cai Y. and Young B. Web crippling of lean duplex stainless steel tubular sections under concentrated end bearing loads. *Thin-Walled Structures* 134, 29–39, 2019.

[39] Feng R., Liu Y. and Zhu J. Tests of CHS-to-SHS tubular connections in stainless steel. *Engineering Structures* 199, 109590, 2019.

[40] Gardner L. and Nethercot D.A. Experiments on stainless steel hollow sections – Part 1: Material and cross-sectional behaviour. *Journal of Constructional Steel Research* 60, 1291–1318, 2004.

[41] Gardner L. and Nethercot D.A. Experiments on stainless steel hollow sections – Part 2: Member behaviour of columns and beams. *Journal of Constructional Steel Research* 60, 1319–1332, 2004.

[42] Gardner L., Talja A. and Baddoo N.R. Structural design of high-strength austenitic stainless steel. *Thin-Walled Structures* 44, 517–528, 2006.

[43] Huang Y. and Young B. Material properties of cold-formed lean duplex stainless steel sections. *Thin-Walled Structures* 54, 72–81, 2012.

[44] Huang Y. and Young B. Tests of pin-ended cold-formed lean duplex stainless steel columns. *Journal of Constructional Steel Research* 82, 203–215, 2013.

[45] Huang Y. and Young B. Experimental investigation of cold-formed lean duplex stainless steel beam-columns. *Thin-Walled Structures* 76, 105–117, 2014.

[46] Hyttinen V. Design of cold-formed stainless steel SHS beam-columns. Report 41, University of Oulu, Oulu, Finland, 1994.

[47] Islam S.M.Z. and Young B. Ferritic stainless steel tubular members strengthened with high modulus CFRP plate subjected to web crippling. *Journal of Constructional Steel Research* 77, 107–118, 2012.

[48] Islam S.M.Z. and Young B. Strengthening of ferritic stainless steel tubular structural members using FRP subjected to Two-Flange-Loading. *Thin-Walled Structures* 62, 179–190, 2013.

[49] Islam S.M.Z. and Young B. FRP strengthening of lean duplex stainless steel hollow sections subjected to web crippling. *Thin-Walled Structures* 85, 183–200, 2014.

[50] Li H.T. and Young B. Cold-formed ferritic stainless steel tubular structural members subjected to concentrated bearing loads. *Engineering Structures* 145, 392–405, 2017.

[51] Li H.T. and Young B. Web crippling of cold-formed ferritic stainless steel square and rectangular hollow sections. *Engineering Structures* 176, 968–980, 2018.

[52] Liu Y. and Young B. Buckling of stainless steel square hollow section compression members. *Journal of Constructional Steel Research* 59, 165–177, 2003.

[53] Lui W.M., Ashraf M. and Young B. Tests of cold-formed duplex stainless steel SHS beam–columns. *Engineering Structures* 74, 111–121, 2014.

[54] Rasmussen K.J.R. and Hancock G.J. Design of Cold-Formed Stainless Steel Tubular Members. I: Columns. *Journal of Structural Engineering* 119(8), 2349–2367, 1993.

[55] Real E. and Mirambell E. Flexural behaviour of stainless steel beams. *Engineering Structures* 27, 1465–1475, 2005.

[56] Talja A. and Salmi P. Design of Stainless Steel RHS Beams, Columns and Beam Columns. Research Note 1619, VTT Building Technology, Finland, 1995.

[57] Theofanous M. and Gardner L. Testing and numerical modelling of lean duplex stainless steel hollow section columns. *Engineering Structures* 31, 3047–3058, 2009.

[58] Theofanous M. and Gardner L. Experimental and numerical studies of lean duplex stainless steel beams. *Journal of Constructional Steel Research* 66, 816–825, 2010.

[59] Theofanous M., Saliba N., Zhao O. and Gardner L. Ultimate response of stainless steel continuous beams. *Thin-Walled Structures* 83, 115–127, 2014.

[60] VTT. WP 3 Members with Class 4 cross-sections in fire. Technical Report, VTT Building Technology, Finland, 2007.

[61] Young B. and Liu Y. Experimental Investigation of Cold-Formed Stainless Steel Columns. *Journal of Structural Engineering (ASCE)* 129, 169–176, 2003.

[62] Young B. and Lui W.M. Behavior of Cold-Formed High Strength Stainless Steel Sections. *Journal of Structural Engineering (ASCE)* 131(11), 1738–1745, 2005.

[63] Young B. and Lui W.M. Tests of cold-formed high strength stainless steel compression members. *Thin-Walled Structures* 44, 224–234, 2006.

[64] Zhao O., Rossi B., Gardner L. and Young B. Behaviour of structural stainless steel cross-sections under combined loading – Part I: Experimental study. *Engineering Structures* 89, 236–246, 2015.

[65] Zhao O., Rossi B., Gardner L. and Young B. Experimental and Numerical Studies of Ferritic Stainless Steel Tubular Cross Sections under Combined Compression and Bending. *Journal of Structural Engineering (ASCE),* 04015110-2, 2015.

[66] Zhao O., Gardner L. and Young B. Buckling of ferritic stainless steel members under combined axial compression and bending. *Journal of Constructional Steel Research* 117, 35–48, 2016.

[67] Zheng B., Hua X. and Shu G. Tests of cold-formed and welded stainless steel beam-columns. *Journal of Constructional Steel Research* 111, 1–10, 2015.

[68] Zhou F. and Young B. Tests of cold-formed stainless steel tubular flexural members. *Thin-Walled Structures* 43, 1325–1337, 2005.

[69] Zhou F. and Young B. Design and tests of cold-formed stainless steel sections subjected to concentrated bearing load. *Advances in Steel Structures* 1, 487–496, 2005.

[70] Zhou F. and Young B. Cold-Formed Stainless Steel Sections Subjected to Web Crippling. *Journal of Structural Engineering (ASCE)* 132(1), 134–144, 2006.

[71] Zhou F. and Young B. Experimental and numerical investigations of cold-formed stainless steel tubular sections subjected to concentrated bearing load. *Journal of Constructional Steel Research* 63, 1452–1466, 2007.

[72] Bardi F.C. and Kyriakides S. Plastic buckling of circular tubes under axial compression – Part I: Experiments. *International Journal of Mechanical Sciences* 48 830–841, 2006.

[73] Buchanan C., Real E. and Gardner L. Testing, simulation and design of cold-formed stainless steel CHS columns. *Thin-Walled Structures* 130, 297–312, 2018.

[74] Burgan B.A., Baddoo N.R. and Gilsenan K.A. Structural design of stainless steel members – comparison between Eurocode 3, Part 1.4 and test results. *Journal of Constructional Steel Research* 54, 51–73, 2000.

[75] He A. and Zhao O. Experimental and numerical investigations of concrete-filled stainless steel tube stub columns under axial partial compression. *Journal of Constructional Steel Research* 158, 405–416, 2019.

[76] Kiymaz G. Strength and stability criteria for thin-walled stainless steel circular hollow section members under bending. *Thin-Walled Structures* 43, 1534–1549, 2005.

[77] Rasmussen K.J.R. and Hancock J.G. Design of Cold-Formed Stainless Steel Tubular Members. II: Beams. *Journal of Structural Engineering* 119(8), 2368–2386, 1993.

[78] Talja A. Test report on welded I and CHS beams, columns and beam-columns. Research Report for WP3. VTT Building Technology, Finland, 1997.

[79] Young B. and Hartono W. Compression Tests of Stainless Steel Tubular Members. *Journal of Structural Engineering (ASCE)* 128,754–761, 2002.

[80] Zhao O., Gardner L. and Young B. Structural performance of stainless steel circular hollow sections under combined axial load and bending – Part1: Experiments and numerical modelling. *Thin-Walled Structures* 101, 231–239, 2016.

[81] Zhao O., Gardner L. and Young B. Testing and numerical modelling of austenitic stainless steel CHS beam–columns. *Engineering Structures* 111, 263–274, 2016.

[82] Bu Y. and Gardner L. Laser-welded stainless steel I-section beam-columns: Testing, simulation and Design. *Engineering Structures* 179, 23–36, 2019.

[83] dos Santos, G.B. and Gardner L. Testing and numerical analysis of stainless steel I-sections under concentrated end-one-flange loading. *Journal of Constructional Steel Research* 157, 271–281, 2019.

[84] dos Santos, G.B., Gardner L. and Kucuckler M. Experimental and numerical study of stainless steel I-sections under concentrated internal one-flange and internal two-flange loading. *Engineering Structures* 175, 355–370, 2018.

[85] Gardner L., Bu Y. and Theofanous M. Laser-welded stainless steel I-sections: Residual stress measurements and column buckling tests. *Engineering Structures* 127, 536–548, 2016.

[86] Sun Y. and Zhao O. Material response and local stability of high-chromium stainless steel welded I-sections. *Engineering Structures* 178, 212–226, 2019.

[87] Wang Y., Yang L., Gao B., Shi Y. and Yuan H. Experimental Study of Lateral-torsional Buckling Behavior of Stainless Steel Welded I-section Beams. *International Journal of Steel Structures* 14(2), 411–420, 2014.

[88] Liang Y., Zhao O., Long Y.L. and Gardner L. Stainless steel channel sections under combined compression and minor axis bending – Part 1: Experimental study and numerical modelling. *Journal of Constructional Steel Research* 152, 154–161, 2019.

[89] Theofanous M., Liew A. and Gardner L. Experimental study of stainless steel angles and channels in bending. *Structures* 4, 80–90, 2015.

[90] Yousefi A.M., Lim J.B.P. and Clifton G.C. Web bearing capacity of unlipped cold-formed ferritic stainless steel channels with perforated web subject to end-two-flange (ETF) loading. *Engineering Structures* 152, 804–818, 2017.

[91] Yousefi A.M., Lim J.B.P. and Clifton G.C. Cold-formed ferritic stainless steel unlipped channels with web openings subjected to web crippling under interior-two-flange loading condition – Part I: Tests and finite element model validation. *Thin-Walled Structures* 116, 333–341, 2017.

[92] Yousefi A.M., Lim J.B.P. and Clifton G.C. Web Crippling Behavior of Unlipped Cold-Formed Ferritic Stainless Steel Channels Subject to One-Flange Loadings. *Journal of Structural Engineering (ASCE)* 144(8), 04018105, 2018.

[93] Yousefi A.M., Lim J.B.P. and Clifton G.C. Cold-formed ferritic stainless steel unlipped channels with web perforations subject to web crippling under one-flange loadings. *Construction and Building Materials* 191, 713–725, 2018.

[94] Yousefi A.M., Lim J.B.P. and Clifton G.C. Web crippling strength of perforated cold-formed ferritic stainless steel unlipped channels with restrained flanges under one-flange loadings. *Thin-Walled Structures* 137, 94–105, 2019.

[95] Yousefi A.M., Lim J.B.P. and Clifton G.C. Web crippling design of cold-formed ferritic stainless steel unlipped channels with fastened flanges under end-two-flange loading condition. *Journal of Constructional Steel Research* 152, 12–28, 2019.

[96] Becque J. and Rasmussen K.J.R. Experimental investigation of local-overall interaction buckling of stainless steel lipped channel columns. *Journal of Constructional Steel Research* 65, 1677–1684, 2009.

[97] Lecce M. and Rasmussen K.J.R. Distortional Buckling of Cold-Formed Stainless Steel Sections: Experimental Investigation. *Journal of Structural Engineering (ASCE)* 132(4), 497–504, 2006.

[98] Niu S., Rasmussen K.J.R. and Fan F. Distortional–global interaction buckling of stainless steel C-beams: Part I — Experimental investigation. *Journal of Constructional Steel Research* 96, 127–139, 2014.

[99] Liang Y., Jeyapragasam V.V.K., Zhang L. and Zhao O. Flexural-torsional buckling behaviour of fixed-ended hot-rolled austenitic stainless steel equal-leg angle section columns. *Journal of Constructional Steel Research* 154, 43–54, 2019.

[100] de Menezes A.A., da S. Vellasco P.C.G., de Lima L.R.O., da Silva A.T. Experimental and numerical investigation of austenitic stainless steel hot-rolled angles under compression. *Journal of Constructional Steel Research* 152, 42–56, 2019.

[101] Sun Y., Liu Z., Liang Y. and Zhao O. Experimental and numerical investigations of hot-rolled austenitic stainless steel equal-leg angle sections. *Thin-Walled Structures* 144, 106225, 2019.

[102] Zhang L., Tan K.H. and Zhao O. Experimental and numerical studies of fixed-ended cold-formed stainless steel equal-leg angle section columns. *Engineering Structures* 184, 134–144, 2019.

[103] Steel Construction Institute (SCI). Design Manual for Structural Stainless Steel. Fourth Edition, 2017.

[104] American Institute of Steel Construction (AISC). Design Guide 27: Structural Stainless Steel. Illinois, USA, 2013.

[105] Mirambell E. and Real E. On the calculation of deflections in structural stainless steel beams: an experimental and numerical investigation. *Journal of Constructional Steel Research* 54(4), 109-133, 2000.

[106] Ramberg W. and Osgood W.R. Description of stress-strain curves by three parameters. Technical Note No. 902. Washington, D.C., USA. National Advisory Committee for Aeronautics, 1943.

[107] Hill H.N. Determination of stress-strain relations from "offset" yield strength values. Technical Note No. 927, 1944.

[108] European Committee for Standardization (CEN). EN 10088-1. Stainless Steels Part 1: List of stainless steels. Brussels, Belgium, 2005.

[109] European Committee for Standardization (CEN). EN 10088-4. Stainless Steels Part 4: Technical Delivery Conditions for Sheet/Plate and Strip of Corrosion Resisting Steels for Construction Purposes. Brussels, Belgium, 2009.

[110] Arrayago I., Real E. and Gardner L. Description of stress–strain curves for stainless steel alloys. *Materials and Design* 87, 540–552, 2015.

[111] Afshan S, Zhao O, Gardner L. Standardised material properties for numerical parametric studies of stainless steel structures and buckling curves for tubular columns. *Journal of Constructional Steel Research* 152, 2–11, 2019.

[112] Steel Construction Institute (SCI). Proposed material factors for EN 1993-1-4. Technical Report. UK, 2020.

[113] American Society for Testing, and Materials (ASTM). Standard Specification for Austenitic Stainless Steel, Sheet, Strip, Plate, and Flat Bar for Structural Applications. ASTM Designation: ASTM A666-84, 1984.

[114] Afshan S., Rossi B. and Gardner L. Strength enhancements in cold-formed structural sections – Part I: Material testing. *Journal of Constructional Steel Research* 83, 177–188, 2013.

[115] Becque J., Rasmussen K.J.R. Experimental investigation of the interaction of local and overall buckling of stainless steel I-columns. *Journal of Structural Engineering (ASCE)* 135(11), 1340–1348, 2009.

[116] Elflah M., Theofanous M., Dirar S., Yuan H. Behaviour of stainless steel beam-to-column joints – Part 1: Experimental investigation. *Journal of Constructional Steel Research* 152, 183–193, 2019.

[117] Real E., Mirambell E. and Estrada I. Shear response of stainless steel plate girders. *Engineering Structures* 29, 1626–1640, 2007.

[118] Fan S., Liu F., Zheng B., Shu G. and Tao Y. Experimental study on bearing capacity of stainless steel lipped C section stub columns. *Thin-Walled Structures* 83, 70–84, 2014.

[119] Han L.H., Chen F., Liao F.Y., Tao Z. and Uy B. Fire performance of concrete filled stainless steel tubular columns. *Engineering Structures* 56, 165–181, 2013.

[120] Ning K., Yang L., Yuan H. and Zhao M. Flexural buckling behaviour and design of welded stainless steel box-section beam-columns. *Journal of Constructional Steel Research* 161, 47–56, 2019.

[121] Nip K.H., Gardner L. and Elghazouli A.Y. Cyclic testing and numerical modelling of carbon steel and stainless steel tubular bracing members. *Engineering Structures* 32(2), 424–441, 2010.

[122] Rasmussen K.J.R. Full-range stress–strain curves for stainless steel alloys. Research Report No. R811, Department of Civil Engineering. The University of Sydney, 2001.

[123] Rasmussen K.J.R. and A.S. Hasham A.S. Tests of X- and K-joints in CHS stainless steel tubes. *Journal of Structural Engineering (ASCE)* 127(10), 1183–1189, 2001.

[124] Rasmussen K.J.R. and Young B. Tests of X- and K-joints in SHS stainless steel tubes. *Journal of Structural Engineering (ASCE)* 127(10), 1173–1182, 2001.

[125] Uy B., Tao Z. and Han L.H. Behaviour of short and slender concrete-filled stainless steel tubular columns. *Journal of Constructional Steel Research* 67, 360–378, 2011.

[126] Xu M. and Szalyga M. Comparative Investigations on the Load-bearing Behaviour of Single Lap Joints with Bolts Stressed in Shear and Bearing-experimental and Simulation. Master Project at the University of Duisburg-Essen, Institute for Metal and Lightweight Structures, 2011.

[127] Young B. and Hartono W. Compression Tests of Stainless Steel Tubular Members. *Journal of Structural Engineering (ASCE)* 128, 754–761, 2002.

[128] Yousuf M., Uy B., Tao Z., Remennikov A. and Liew J.Y.R. Transverse impact resistance of hollow and concrete filled stainless steel columns. *Journal of Constructional Steel Research* 82, 177–189, 2013.

[129] Manninen T. Technical report for Structural applications of ferritic stainless steels (SAFSS). WP1 end-user requirements and material performance. Task 1.3 Characterization of stress– strain behaviour: Technical Specifications for Room-Temperature: Tensile and Compression Testing, 2011.

[130] Real E., Arrayago I., Mirambell E. and Westeel R. Comparative study of analytical expressions for the modelling of stainless steel behaviour. *Thin-Walled Structures* 83 2–11, 2014.

[131] Rossi B. Mechanical behaviour of ferritic grade 3Cr12 stainless steel. Part 1: Experimental investigations. *Thin-Walled Structures* 48, 553–560, 2010.

[132] Talja A. and Hradil P. Structural performance of steel members: model calibration tests. Research report for SAFSS-WP2. Internal Report, VTT Building Technology, Finland, 2011.

[133] Tondini N., Rossi B., Franssen J.M. Experimental investigation on ferritic stainless steel columns in fire. *Fire Safety Journal* 62, 238–248, 2013.

[134] Ellobody E. and Young B. Structural performance of cold-formed high strength stainless steel columns. *Journal of Constructional Steel Research* 61, 1631–1649, 2005.

[135] Rasmussen K.J.R., Burns T., Bezkorovainy P., Bambach M.R. Numerical modelling of stainless steel plates in compression. *Journal of Constructional Steel Research* 59, 1345–1362, 2003.

[136] Saliba N. and Gardner L. Experimental study of the shear response of lean duplex stainless steel plate girders. *Engineering Structures* 46, 375–391, 2013.

[137] Saliba N. and Gardner L. Cross-section stability of lean duplex stainless steel welded I-sections. *Journal of Constructional Steel Research* 80, 1–14, 2013.

[138] Young B. and Ellobody E. Experimental investigation of concrete-filled cold-formed high strength stainless steel tube columns. *Journal of Constructional Steel Research* 62, 484–492, 2006.

[139] Australian/New Zealand Standard (AS/NZS). AS/NZS 4673:2001. Cold-formed stainless steel structures. Sydney, Australia, 2001.

[140] American Society of Civil Engineers (ASCE). SEI/ASCE 8-02. Specification for the Design of Cold-Formed Stainless Steel Structural Members. Virginia, USA, 2002.

[141] European Committee for Standardization (CEN). EN ISO 6892-1. Metallic materials - Tensile testing - Part 1: Method of test at room temperature. Brussels, Belgium, 2009.

[142] Australian Standards (AS). AS 1391-2020. Metallic Materials - Tensile Testing at Ambient Temperature. Sydney, Australia, 2020.

[143] American Society for Testing and Materials (ASTM). ASTM E8/E8M. Standard Test Methods for Tension Testing for Metallic Materials. West Conshohocken (USA), 2016.

[144] Rossi B., Afshan S. and Gardner L. Strength enhancements in cold-formed structural sections – Part II: Predictive models. *Journal of Constructional Steel Research* 83, 189–196, 2013.

[145] European Committee for Standardization. EN 1990. European Committee for Standardization Eurocode. Basis of structural design. Brussels, Belgium, 2005.

[146] Dawson R.G. and Walker A.C. Post-buckling of geometrically imperfect plates. *Journal of the Structural Division (ASCE)* 98(1), 75–94, 1972.

[147] Gardner L. and Nethercot D.A. Numerical modelling of stainless steel structural components - A consistent approach. *Journal of Structural Engineering (ASCE)* 130(10), 1586–1601, 2004.

[148] Shayan S., Rasmussen K.J.R. and Zhang Hao. On the modelling of initial geometric imperfections of steel frames in advanced analysis. *Journal of Constructional Steel Research* 98, 167–177, 2014.

[149] Shen Y. and Chacón R. Effect of Uncertainty in Localized Imperfection on the Ultimate Compressive Strength of Cold-Formed Stainless Steel Hollow Sections. *Applied Sciences* 9, 3827, 2019.

[150] Beaulieu D. and Adams P.F. The results of a survey on structural out-of-plumbs. *Canadian Journal of Civil Engineering* 5(4), 462–70, 1978.

[151] Lindner J. and Gietzelt R. Assumptions for imperfections for out-of-plumb of columns. *Stahlbau* 53, 97–102, 1984.

[152] Cruise R.B. and Gardner L. Residual stress analysis of structural stainless steel sections. *Journal of Constructional Steel Research* 64 352–366, 2008.

[153] Gardner L. and Cruise R.B. Modeling of Residual Stresses in Structural Stainless Steel Sections. Journal of Structural Engineering (ASCE) 135(1), 42–53, 2009.

[154] Jandera M., Gardner L. and Machacek J. Residual stresses in cold-rolled stainless steel hollow sections. *Journal of Constructional Steel Research* 64, 1255–1263, 2008.

[155] Zheng B., Shu G. and Jiang Q. Experimental study on residual stresses in cold rolled austenitic stainless steel hollow sections. *Journal of Constructional Steel Research* 152, 94–104, 2019.

[156] Yuan H.X., Wang Y.Q., Shi Y.J. and Gardner L. Residual stress distributions in welded stainless steel sections. *Thin-Walled Structures* 79, 38–51, 2014.

[157] Shayan S., Rasmussen K.J.R. and Zhang, H. Probabilistic modelling of residual stress in advanced analysis of steel structures. *Journal of Constructional Steel Research* 101, 407–414, 2014.

**TABLES**

Table 1. Cross-sectional dimension variations for different stainless steel section types.

| Cross-section type | Random variable | No. specimens | Mean | Std. Dev. | COV | Statistical distribution | References |
|---|---|---|---|---|---|---|---|
| SHS/RHS | Height, H | 1070 | $1.002 \cdot H_n$ | $0.012 \cdot H_n$ | 0.012 | Normal | [32-71] |
| | Width, B | 1070 | $1.002 \cdot B_n$ | $0.010 \cdot B_n$ | 0.010 | Normal | |
| | Thickness, t | 1058 | $0.974 \cdot t_n$ | $0.028 \cdot t_n$ | 0.029 | Normal | |
| | Int. Radius, R | 842 | $1.030 \cdot t_n$ | $0.317 \cdot t_n$ | 0.308 | - | |
| | Ext. Radius, R | 181 | $2.180 \cdot t_n$ | $0.307 \cdot t_n$ | 0.141 | - | |
| CHS | Diameter, D | 165 | $0.999 \cdot D_n$ | $0.003 \cdot D_n$ | 0.003 | Normal | [39,54,72-81] |
| | Thickness, t | 162 | $0.982 \cdot t_n$ | $0.035 \cdot t_n$ | 0.035 | Normal | |
| I-sections | Height, H | 204 | $0.998 \cdot H_n$ | $0.007 \cdot H_n$ | 0.007 | Normal | [55,59,67,74, 78,82-87] |
| | Width, B | 204 | $1.001 \cdot B_n$ | $0.007 \cdot B_n$ | 0.007 | Normal | |
| | Web thickness, $t_w$ | 165 | $0.999 \cdot t_n$ | $0.027 \cdot t_n$ | 0.027 | Normal | |
| | Flange thickness, $t_f$ | 172 | $1.000 \cdot t_n$ | $0.031 \cdot t_n$ | 0.031 | Normal | |
| Channels – Welded | Height, H | 27 | $1.003 \cdot H_n$ | $0.003 \cdot H_n$ | 0.003 | Normal | [88-89] |
| | Width, B | 27 | $0.995 \cdot B_n$ | $0.005 \cdot B_n$ | 0.005 | Normal | |
| | Thickness, t | 27 | $0.975 \cdot t_n$ | $0.033 \cdot t_n$ | 0.034 | Normal | |
| Channels – Cold-formed | Height, H | 260 | $1.008 \cdot H_n$ | $0.003 \cdot H_n$ | 0.003 | Normal | [90-98] |
| | Width, B | 260 | $1.001 \cdot B_n$ | $0.004 \cdot B_n$ | 0.004 | Normal | |
| | Lip, C | 80 | $0.988 \cdot C_n$ | $0.018 \cdot C_n$ | 0.018 | Normal | |
| | Thickness, t | 260 | $0.965 \cdot t_n$ | $0.026 \cdot t_n$ | 0.027 | Normal | |
| Angles – Hot-rolled | Leg, B | 49 | $0.998 \cdot B_n$ | $0.008 \cdot B_n$ | 0.008 | Normal | [99-101] |
| | Thickness, t | 49 | $0.977 \cdot t_n$ | $0.027 \cdot t_n$ | 0.028 | Normal | |
| Angles – Cold-formed | Leg, B | 20 | $1.000 \cdot B_n$ | $0.004 \cdot B_n$ | 0.004 | Normal | [89,102] |
| | Thickness, t | 20 | $0.988 \cdot t_n$ | $0.021 \cdot t_n$ | 0.021 | Normal | |

Note: the subscript *n* represents the nominal value.

Table 2. Calculated values for the COV of geometric properties for stainless steel sections in compression and bending.

| Cross-section type | No. specimens | $V_{geom}$ compression | $V_{geom}$ bending |
|---|---|---|---|
| SHS and RHS | 1058 | 0.029 | 0.027 |
| CHS | 162 | 0.035 | 0.035 |
| I-sections | 165 | 0.019 | 0.025 |
| Channels – Cold-formed | 260 | 0.042 | 0.043 |
| Channels – Welded | 27 | 0.023 | 0.023 |
| Angles | 69 | 0.018 | 0.018 |

Table 3. Yield stress $f_y$ and ultimate strength $f_u$ variation proposals in the literature.

| Variable | Stainless steel type | Mean | COV | Reference |
|---|---|---|---|---|
| Yield stress $f_y$ | Austenitic and Ferritic | $1.10^1$ | 0.100 | [18] |
| | Austenitic | $1.30^2$ | 0.060 | |
| | Ferritic | $1.20^2$ | 0.045 | [17] |
| | Duplex | $1.10^2$ | 0.030 | |
| | Austenitic | $1.20^2$ | 0.059 | |
| | Ferritic | $1.15^2$ | 0.056 | [112] |
| | Duplex | $1.10^2$ | 0.054 | |
| Ultimate strength $f_u$ | Austenitic and Ferritic | $1.10^1$ | 0.050 | [18] |
| | All | $1.10^2$ | 0.035 | [17] |
| | All | $1.10^2$ | 0.025 | [112] |

[1]: overstrength ratios correspond to nominal values in ASTM A666-84 [113].
[2]: overstrength ratios correspond to nominal values in EN 10088-4 [109].

Table 4. Young's modulus E variations for different stainless steel and product types.

| Stainless steel type | Product type | No. specimens | Mean [MPa] | Std. Dev. [MPa] | COV | Statistical distribution |
|---|---|---|---|---|---|---|
| Austenitic | Plates & Sheets | 299 | 195416 | 11175 | 0.057 | |
| | Cold-formed | 497 | 193084 | 11039 | 0.057 | Normal |
| | All | 796 | 193960 | 11141 | 0.057 | |
| Ferritic | Plates & Sheets | 93 | 203738 | 14764 | 0.072 | |
| | Cold-formed | 193 | 198572 | 15582 | 0.078 | Normal |
| | All | 286 | 200252 | 15486 | 0.077 | |
| Duplex and Lean Duplex | Plates & Sheets | 164 | 206233 | 11957 | 0.058 | |
| | Cold-formed | 128 | 205025 | 8700 | 0.042 | Normal |
| | All | 292 | 205703 | 10653 | 0.052 | |

Table 5. Yield stress $f_y$ variations for different stainless steel plates and sheet materials.

| Stainless steel type | No. specimens | Mean | Std. Dev. | COV | Statistical distribution |
|---|---|---|---|---|---|
| Austenitic | 306 | $1.22 \cdot f_{y,n}$ | $0.075 \cdot f_{y,n}$ | 0.061 | Log-normal |
| Ferritic | 72 | $1.22 \cdot f_{y,n}$ | $0.063 \cdot f_{y,n}$ | 0.052 | Log-normal |
| Duplex and Lean Duplex | 83 | $1.11 \cdot f_{y,n}$ | $0.097 \cdot f_{y,n}$ | 0.088 | Log-normal |

Note: the subscript *n* represents the nominal value in EN 10088-4 [109].

Table 6. Yield stress $f_y$ variations for different stainless steel cold-formed materials.

| Stainless steel type | Reference yield stress | No. specimens | Mean | Std. Dev. | COV | Statistical distribution |
|---|---|---|---|---|---|---|
| Austenitic | Nominal | 363 | $1.78 \cdot f_{y,n}$ | $0.353 \cdot f_{y,n}$ | 0.199 | Log-normal |
|  | Enhanced nominal |  | $1.39 \cdot f_{y,enh,n}$ | $0.220 \cdot f_{y,enh,n}$ | 0.158 | Log-normal |
| Ferritic | Nominal | 175 | $1.60 \cdot f_{y,n}$ | $0.257 \cdot f_{y,n}$ | 0.161 | Log-normal |
|  | Enhanced nominal |  | $1.38 \cdot f_{y,enh,n}$ | $0.163 \cdot f_{y,enh,n}$ | 0.118 | Log-normal |
| Duplex and Lean Duplex | Nominal | 122 | $1.18 \cdot f_{y,n}$ | $0.166 \cdot f_{y,n}$ | 0.140 | Log-normal |
|  | Enhanced nominal |  | $1.16 \cdot f_{y,enh,n}$ | $0.140 \cdot f_{y,enh,n}$ | 0.121 | Log-normal |

Note: the subscripts *n* and *enh_n* represent the nominal value in EN 10088-4 [109] and the enhanced nominal values from [103,144], respectively.

Table 7. Ultimate strength $f_u$ variations for different stainless steel and product types.

| Stainless steel type | Product type | No. specimens | Mean | Std. Dev. | COV | Statistical distribution |
|---|---|---|---|---|---|---|
| Austenitic | Plates & Sheets | 219 | $1.15 \cdot f_{u,n}$ | $0.040 \cdot f_{u,n}$ | 0.034 | Log-normal |
|  | Cold-formed | 274 | $1.33 \cdot f_{u,n}$ | $0.124 \cdot f_{u,n}$ | 0.094 | Log-normal |
| Ferritic | Plates & Sheets | 100 | $1.12 \cdot f_{u,n}$ | $0.099 \cdot f_{u,n}$ | 0.088 | Log-normal |
|  | Cold-formed | 155 | $1.16 \cdot f_{u,n}$ | $0.130 \cdot f_{u,n}$ | 0.112 | Log-normal |
| Duplex and Lean Duplex | Plates & Sheets | 41 | $1.10 \cdot f_{u,n}$ | $0.026 \cdot f_{u,n}$ | 0.023 | Log-normal |
|  | Cold-formed | 86 | $1.17 \cdot f_{u,n}$ | $0.123 \cdot f_{u,n}$ | 0.105 | Log-normal |

Note: the subscript *n* represents the nominal value in EN 10088-4 [109].

Table 8. Ultimate strain $\varepsilon_u$ variations for different stainless steel and product types.

| Stainless steel type | Product type | No. specimens | Mean [mm/mm] | Std. Dev. [mm/mm] | COV | Statistical distribution |
|---|---|---|---|---|---|---|
| Austenitic | Plates & Sheets | 47 | 0.49 | 0.06 | 0.118 | Normal |
|  | Cold-formed | 102 | 0.39 | 0.15 | 0.373 | Normal |
| Ferritic | Plates & Sheets | 98 | 0.17 | 0.05 | 0.280 | Normal |
|  | Cold-formed | 159 | 0.13 | 0.11 | 0.852 | Normal |
| Duplex and Lean Duplex | Plates & Sheets | 85 | 0.29 | 0.05 | 0.171 | Normal |
|  | Cold-formed | 99 | 0.24 | 0.14 | 0.579 | Normal |

Table 9. Strain hardening parameter *n* variations for different stainless steel and product types.

| Stainless steel type | Product type | No. specimens | Mean [-] | Std. Dev. [-] | COV | Statistical distribution |
|---|---|---|---|---|---|---|
| Austenitic | Plates & Sheets | 50 | 10.6 | 1.8 | 0.168 | Log-normal |
| | Cold-formed | 284 | 6.0 | 2.1 | 0.342 | Log-normal |
| Ferritic | Plates & Sheets | 89 | 16.3 | 4.9 | 0.302 | Log-normal |
| | Cold-formed | 188 | 10.1 | 5.4 | 0.536 | Log-normal |
| Duplex and Lean Duplex | Plates & Sheets | 162 | 6.6 | 2.4 | 0.364 | Log-normal |
| | Cold-formed | 120 | 6.3 | 2.1 | 0.332 | Log-normal |

Table 10. Strain hardening parameter *m* variations for different stainless steel and product types.

| Stainless steel type | Product type | No. specimens | Mean [-] | Std. Dev. [-] | COV | Statistical distribution |
|---|---|---|---|---|---|---|
| Austenitic | Plates & Sheets | 47 | 2.3 | 0.3 | 0.144 | Log-normal |
| | Cold-formed | 96 | 3.1 | 1.3 | 0.426 | Log-normal |
| Ferritic | Plates & Sheets | 95 | 2.9 | 0.3 | 0.104 | Log-normal |
| | Cold-formed | 55 | 3.5 | 2.2 | 0.619 | Log-normal |
| Duplex and Lean Duplex | Plates & Sheets | 31 | 3.6 | 0.5 | 0.138 | Log-normal |
| | Cold-formed | 33 | 4.2 | 1.9 | 0.462 | Log-normal |

Table 11. Local imperfection amplitude variations $\gamma_1$ for different stainless steel welded section types.

| Cross-section type | No. specimens | Mean | Std. Dev. | COV | Statistical distribution | References |
|---|---|---|---|---|---|---|
| I-sections | 59 | $0.0012 \cdot B_n$ | $0.0013 \cdot B_n$ | 1.041 | Normal | [83-86] |
| Channels | 27 | $0.007 \cdot H_n$ | $0.0038 \cdot H_n$ | 0.541 | - | [88-89] |
| Angles | 40 | $0.018 \cdot B_n$ | $0.0009 \cdot B_n$ | 0.515 | Log-normal | [99-101] |

Table 12. Local imperfection amplitude variations $\gamma_2$ for different stainless steel cold-formed section types.

| Cross-section type | No. specimens | Mean | Std. Dev. | COV | Statistical distribution | References |
|---|---|---|---|---|---|---|
| SHS and RHS – Group 1 | 160 | 0.041 | 0.026 | 0.636 | Log-normal | [32-37,42,57,58, 64-66,147] |
| SHS and RHS – Group 2 | 104 | 0.117 | 0.057 | 0.484 | Log-normal | [43,47,49,62] |
| Channels | 16 | 0.036 | 0.030 | 0.828 | Log-normal | [97] |
| Angles | 16 | 0.055 | 0.064 | 1.161 | Log-normal | [102] |

Table 13. Midspan member imperfection amplitude variations for different stainless steel section types.

| Cross-section type | No. specimens | Mean | Std. Dev. | COV | Statistical distribution | References |
|---|---|---|---|---|---|---|
| SHS and RHS | 226 | L/3232 | L/5369 | 0.602 | Log-normal | [32,35,36,41,44-46,52-54,57,61,62,66,67] |
| CHS | 55 | L/3484 | L/6056 | 0.575 | Log-normal | [54,73,78,79,81] |
| I-sections | 43 | L/4038 | L/7013 | 0.576 | Log-normal | [67,82,85,87] |
| Channels | 29 | L/5879 | L/12016 | 0.489 | Log-normal | [96] |
| Angles | 16 | L/4784 | L/25348 | 0.189 | Log-normal | [99] |

Table 14. Member imperfection scale factor variations for different stainless steel section types.

| Cross-section type | No. specimens | Scale factor | Mean | Std. Dev. | Statistical distribution | References |
|---|---|---|---|---|---|---|
| SHS and RHS | 20 | $a_1$ | 0.000300·L | 0.00026·L | Log-normal | [35,36,120] |
|  |  | $a_2$ | 0.000005·L | 0.00014·L | Normal |  |
|  |  | $a_3$ | -0.000010·L | 0.00011·L | Normal |  |
| CHS | 10 | $a_1$ | 0.00026·L | 0.00016·L | Log-normal | [78] |
|  |  | $a_2$ | -0.00001·L | 0.00008·L | Normal |  |
|  |  | $a_3$ | -0.00003·L | 0.00009·L | Normal |  |
| Channels | 60 | $a_1$ | 0.00017·L | 0.00017·L | Log-normal | [96] |
|  |  | $a_2$ | -0.00001·L | 0.00006·L | Normal |  |
|  |  | $a_3$ | 0.00001·L | 0.00004·L | Normal |  |

Table 15. Out-of-plumb angle variations for stainless steel frame imperfections [36].

| Direction | No. specimens | Mean | Std. Dev. |
|---|---|---|---|
| In-plane | 4 | 1/672 | 1/656 |
| Out-of-plane | 4 | 1/159 | 1/531 |

Table 16. Variations of residual stress magnitude scale factors for different stainless steel section types.

| Cross-section type and fabrication | Stainless steel type | RS model | No. specimens | Mean | Std. Dev. | COV | Statistical distribution | References |
|---|---|---|---|---|---|---|---|---|
| Cold-formed SHS and RHS | Austenitic | [153] | 13 | 0.875 | 0.161 | 0.184 | Normal | [152,154,155] |
| Press-braked angles | Austenitic | [153] | 8 | 0.284 | 0.166 | 0.585 | Normal | [152] |
| Welded I-sections | Austenitic and Duplex | [156] | 14 | 0.819 | 0.119 | 0.145 | Normal | [87,156] |
| Welded SHS and RHS | Austenitic and Duplex | [156] | 8 | 0.615 | 0.115 | 0.186 | Normal | [156] |

**FIGURES**

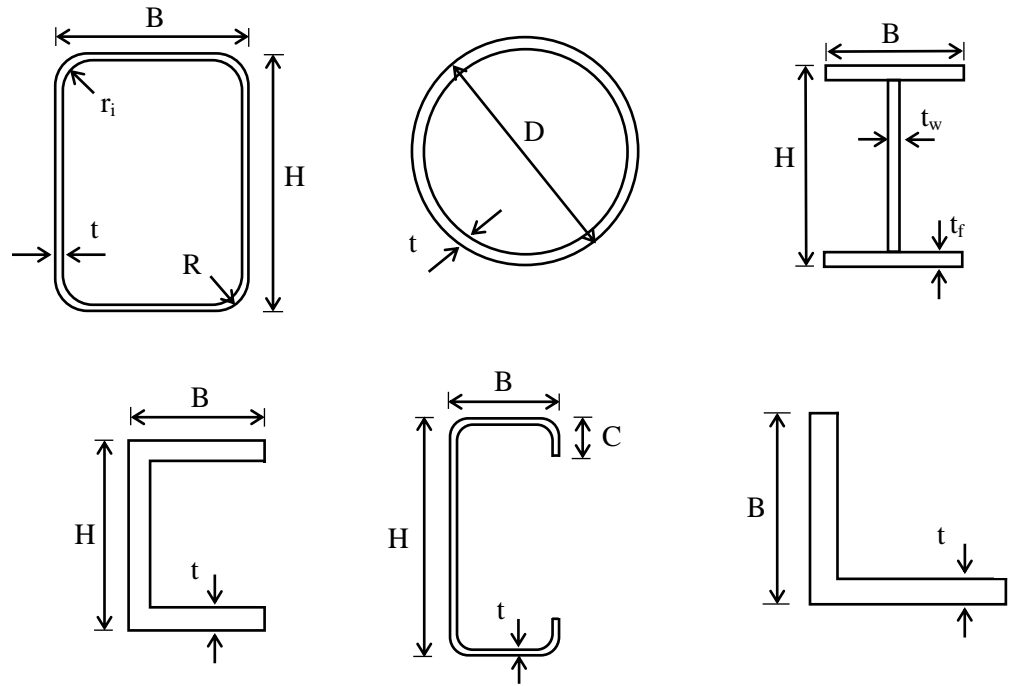

Figure 1. Variables on cross-section properties for different section types.

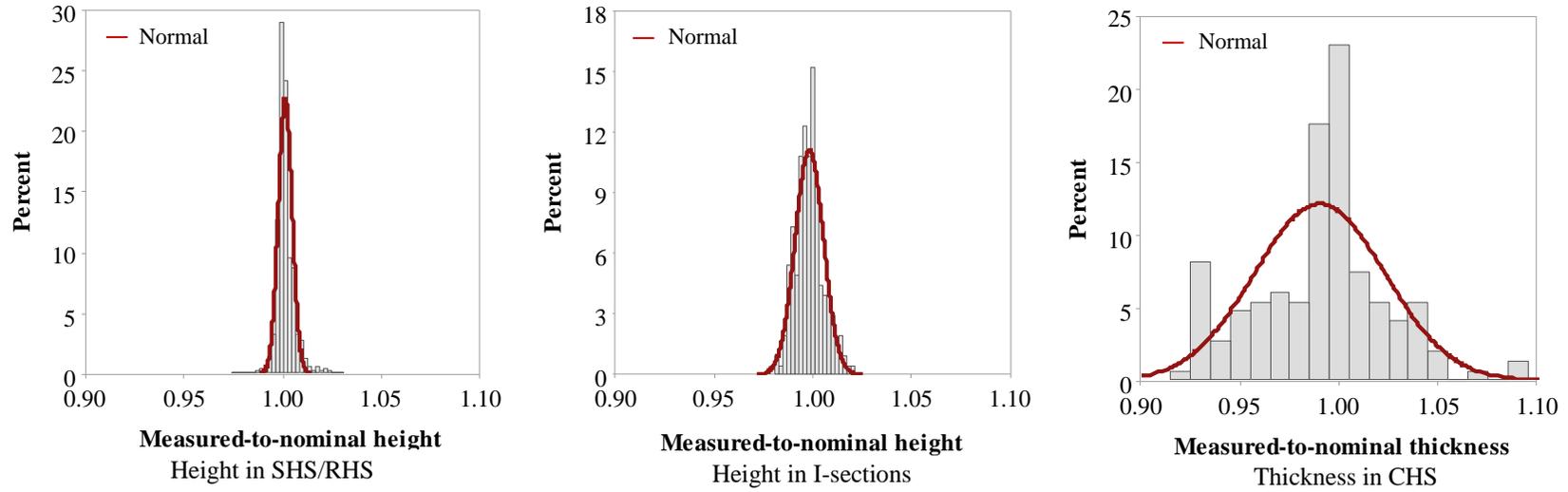

Figure 2. Histograms of cross-section dimensions for different stainless steel section types.

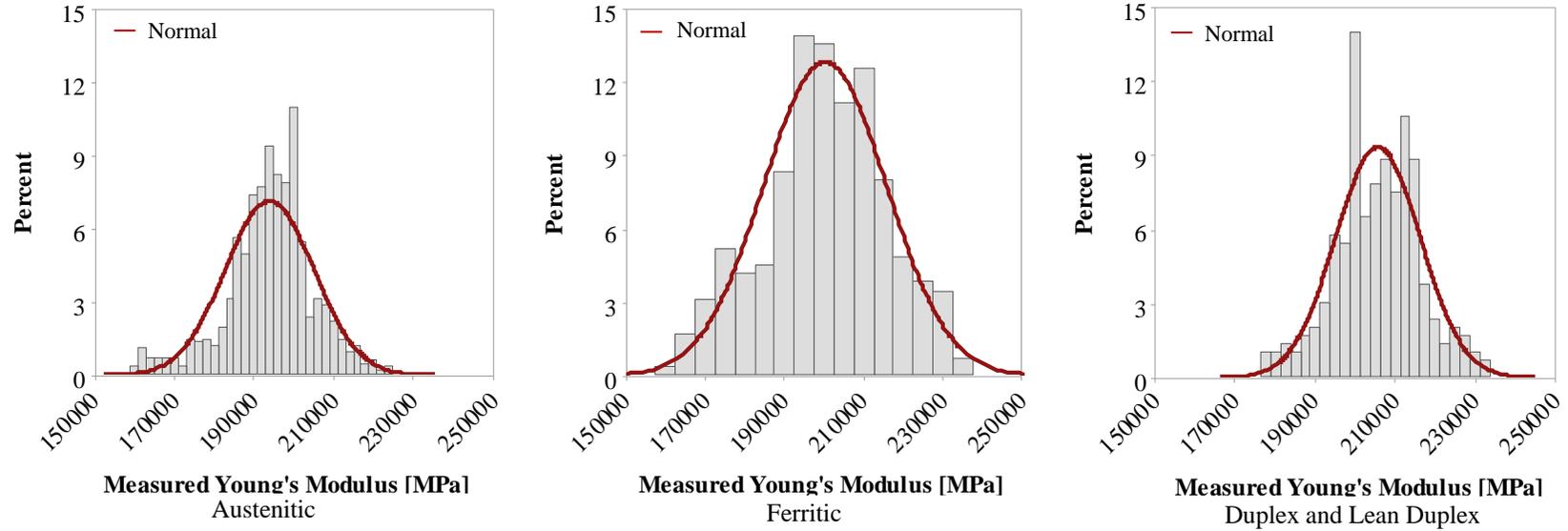

Figure 3. Histograms of Young's modulus E for different stainless steel types.

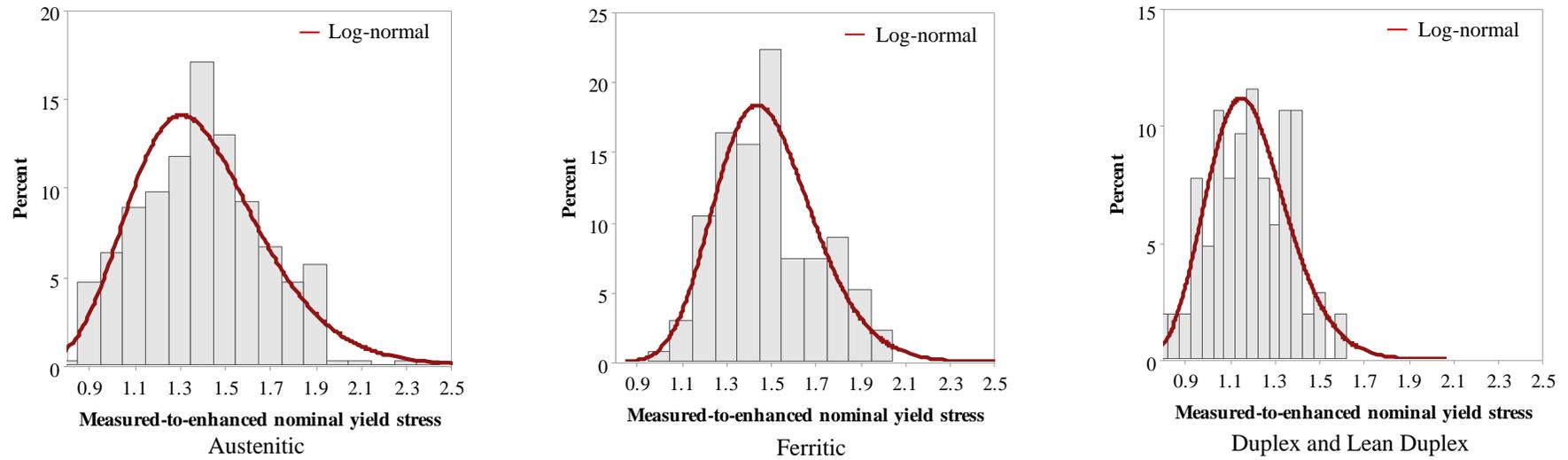

Figure 4. Histograms of yield stress $f_y$ for different cold-formed stainless steel types.

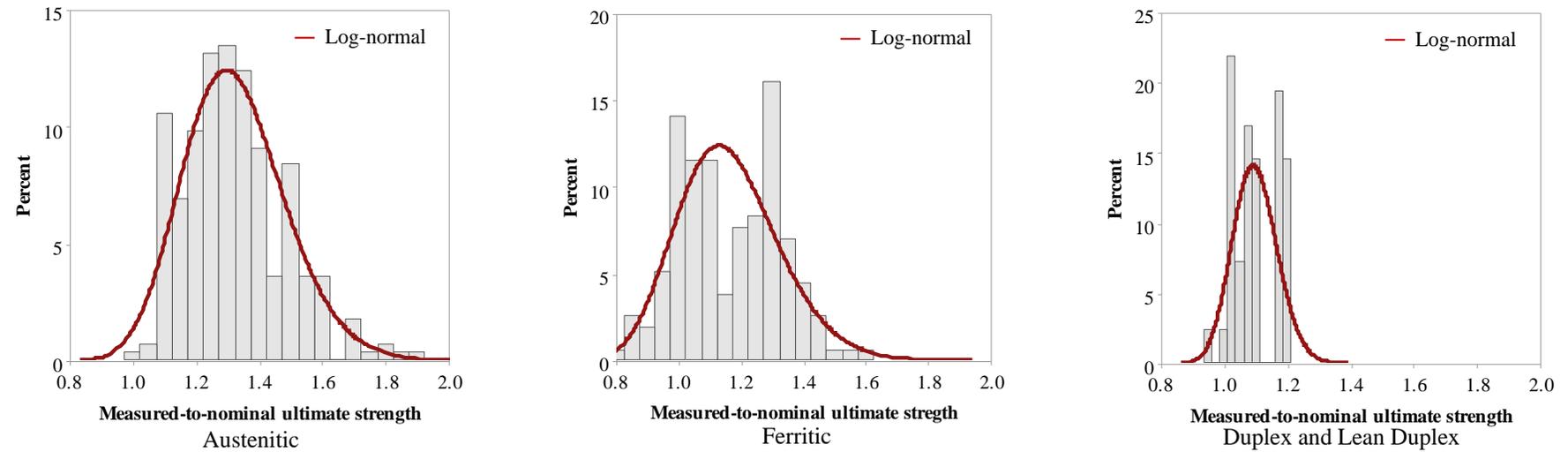

Figure 5. Histograms of ultimate strength $f_u$ for different cold-formed stainless steel types.

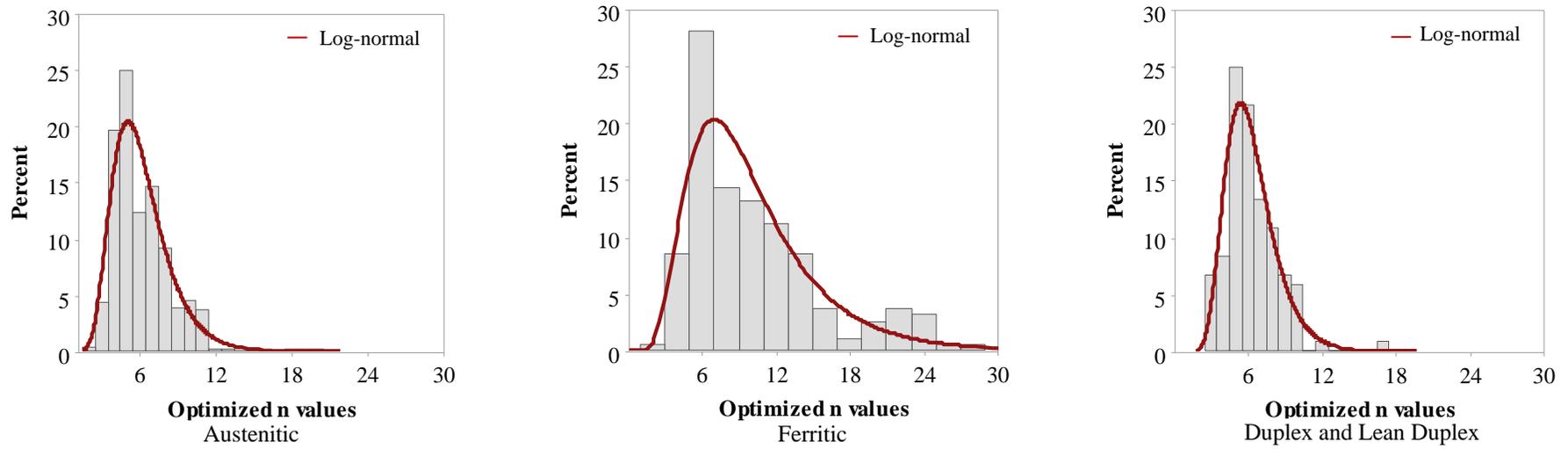

Figure 6. Histograms of strain hardening exponent *n* for different cold-formed stainless steel types.

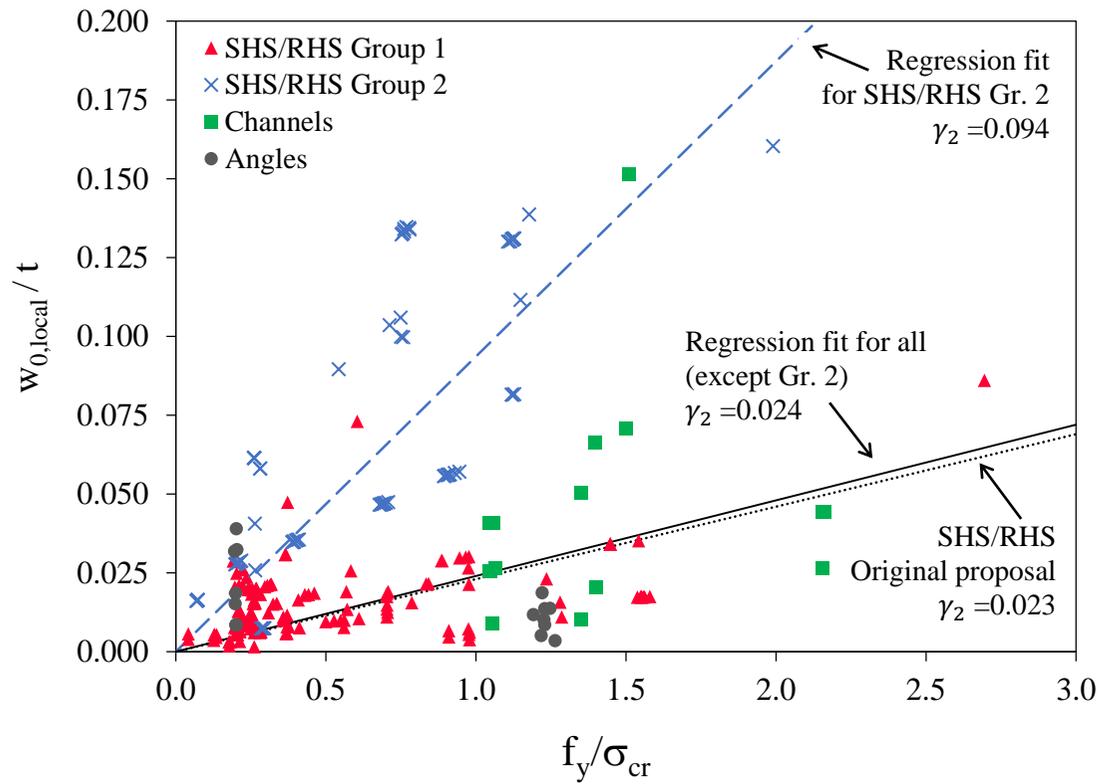

Figure 7. Assessment of predictive models for local imperfection amplitude in stainless steel cold-formed sections.

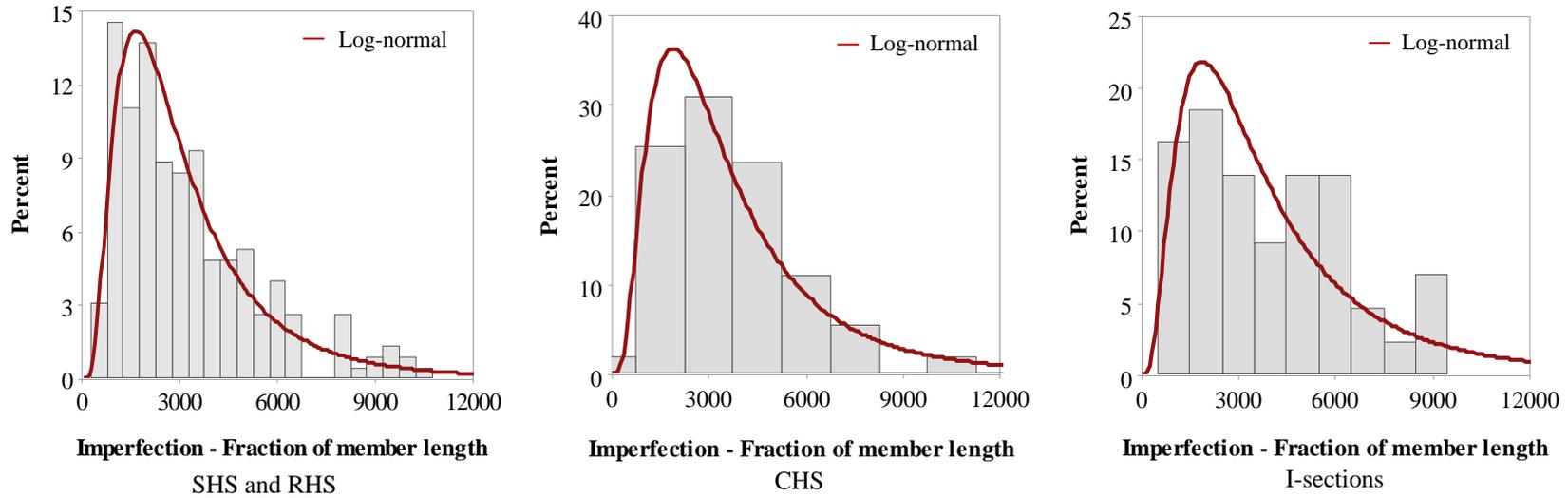
Figure 8. Histograms of member imperfection amplitudes for different stainless steel section types.

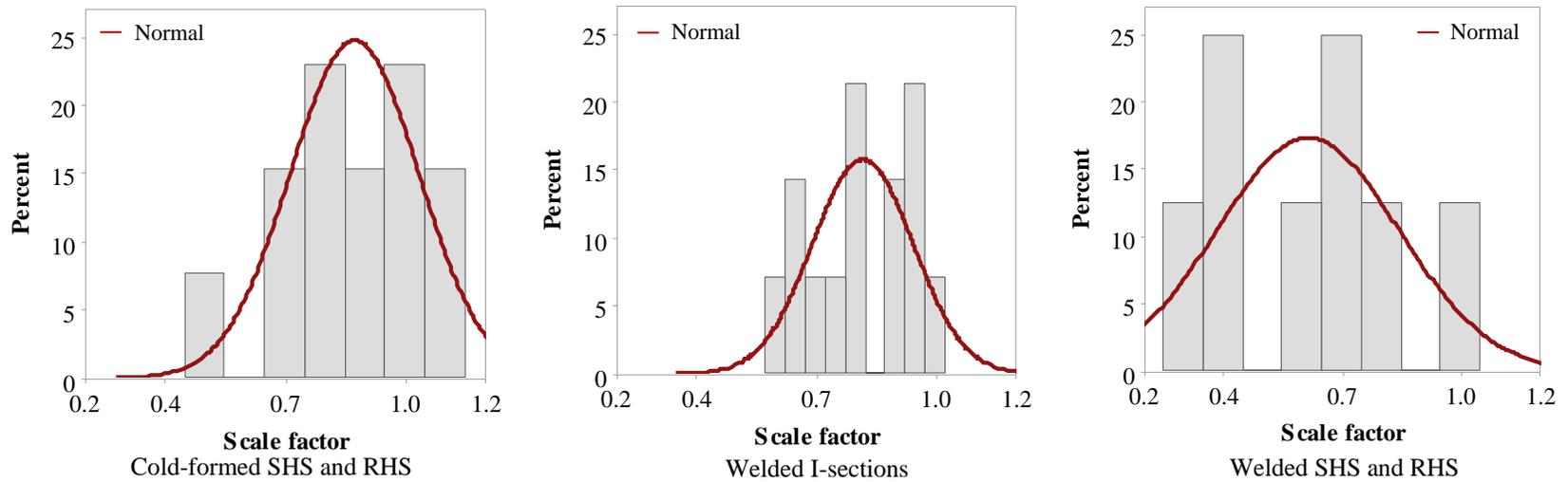
Figure 9. Histograms of residual stress magnitude scale factors for different stainless steel section types.